\documentclass[showpacs,aps,floatfix,prd,nofootinbib]{revtex4}
\usepackage{graphicx}
\usepackage{epsfig}
\usepackage{array, longtable}
\usepackage{amsmath}
\usepackage{colordvi}
\usepackage{hhline}
\newcommand{\BABARPubYear}    {08}

\newcommand{\BABARConfNumber} {016}
\newcommand{\SLACPubNumber} {0000}

\input pubboard/babarsym
\newcommand{\bei}{\begin{itemize}}
\newcommand{\eei}{\end{itemize}}
\newcommand{\beq}{\begin{equation}}
\newcommand{\eeq}{\end{equation}}
\newcommand{\beqn}{\begin{eqnarray}}
\newcommand{\eeqn}{\end{eqnarray}}
\newcommand{\beqns}{\begin{eqnarray*}}
\newcommand{\eeqns}{\end{eqnarray*}}

\newcommand{\equaref}[1]{Eq.~(\ref{eq:#1})}
\newcommand{\equtworef}[2]{Eq.~(\ref{eq:#1}) and (\ref{eq:#2})}

\renewcommand{\figref}[1]{Fig.~\ref{fig:#1}}

\def\ea{{\em et al.}}
\def\exp{{\rm exp}}

\def\min{{\rm min}}
\def\max{{\rm max}}

\def\rPTbarkappa{\kern 0.18em\overline{\kern -0.18em r}{}^{\kappa}{}}
\def\rPTbarsigma{\kern 0.18em\overline{\kern -0.18em r}{}^{\sigma}{}}

\def\deltabarkappa{\kern 0.18em\overline{\kern -0.18em \delta}{}_r^{\kappa}}
\def\deltabarsigma{\kern 0.18em\overline{\kern -0.18em \delta}{}_r^{\sigma}}
\def\deltaTbarkappa{\kern 0.18em\overline{\kern -0.18em \delta}{}_T^{\kappa}}
\def\deltaTbarsigma{\kern 0.18em\overline{\kern -0.18em \delta}{}_T^{\sigma}}

\def\OC{X}
\def\OCbar{{\kern 0.18em\overline{\kern -0.18em \OC}}}

\def\mprime{\ensuremath{m^\prime}}
\def\thetaprime{\ensuremath{\theta^\prime}}
\def\deprime{\ensuremath{{\de^\prime}{}}}

\def\de{\DeltaE}

\def\Qtagi{q_{{\rm tag},i}}

\def\Atagqq{A_{q\bar q,\,\rm tag}}

\def\Atagj{A_{{B,\,\rm tag},j}}

\def\cat{c}

\def\a{\kappa}

\def\Amptpbar{\kern 0.18em\overline{\kern -0.18em {\cal A}}_{{\overline B^0} \rightarrow K^-\pi^+\pi^0}}

\def\Amptpbarkappa{\kern 0.18em\overline{\kern -0.18em A}{}^{\kappa}{}}
\def\Amptpbarsigma{\kern 0.18em\overline{\kern -0.18em A}{}^{\sigma}{}}

\def\Tbarkappa{\kern 0.18em\overline{\kern -0.18em T}{}^{\kappa}{}}
\def\Tbarsigma{\kern 0.18em\overline{\kern -0.18em T}{}^{\sigma}{}}
\def\Pbarkappa{\kern 0.18em\overline{\kern -0.18em P}{}^{\kappa}{}}
\def\Pbarsigma{\kern 0.18em\overline{\kern -0.18em P}{}^{\sigma}{}}

\def\Nbpm{{\kern 0.18em\overline{\kern -0.18em N}}^{+-}}
\def\Nbmp{{\kern 0.18em\overline{\kern -0.18em N}}^{-+}}

\def\Mu{\mu}
\def\Chi2MinaMu{\chi^2_{\min ;\a,\Mu}}
\def\Chi2MinMu{\chi^2_{\min ;\Mu}(a)}

\def\TM{{\rm TM}}
\def\SCF{{\rm SCF}}

\def\fscfave{\kern 0.18em\overline{\kern -0.18em f}_{\rm SCF}}

\def\abar{\bar{a}}

\def\Bbar{\kern 0.18em\overline{\kern -0.18em B}{}\xspace}

\def\BRpmb{{\cal \kern 0.18em\overline{\kern -0.18em  B}}{}_{\rho\pi}^{+-}}
\def\BRmpb{{\cal \kern 0.18em\overline{\kern -0.18em  B}}{}_{\rho\pi}^{-+}}

\def\BRipmb{{\cal \kern 0.18em\overline{\kern -0.18em  B}}{}_{\rho^+\pi^-}}
\def\BRimpb{{\cal \kern 0.18em\overline{\kern -0.18em  B}}{}_{\rho^-\pi^+}}

\def\Abar{\kern 0.18em\overline{\kern -0.18em A}{}}
\def\abar{\kern 0.18em\overline{\kern -0.18em a}{}}

\long\def\inst#1{\par\nobreak\kern 4pt\nobreak
    {\it #1}\par\vskip 10pt plus 3pt minus 3pt}

\begin{document}
{\pagestyle{empty}

\begin{flushright}
\babar-CONF-\BABARPubYear/\BABARConfNumber \\
SLAC-PUB-\SLACPubNumber \\
July 2008 \\
\end{flushright}

\par\vskip 5cm

\begin{center}
\Large \bf Amplitude Analysis of the Decay~$\Bz \to \Kp\pim\piz$
\end{center}
\bigskip

\begin{center}
\large The \babar\ Collaboration\\
\mbox{ }\\
\today
\end{center}
\bigskip \bigskip

\begin{center}
\large \bf Abstract
\end{center}
We report an updated amplitude analysis of the charmless hadronic decays of
neutral \B mesons to $\Kp\pim\piz$. With a sample of
454 million $\FourS \to \B\Bbar$ decays collected by the \babar\ detector 
at the \pep2\ asymmetric-energy \B~Factory at SLAC, we measure the
magnitudes and phases of the intermediate resonant and nonresonant
amplitudes for $\Bz$ and $\Bzb$ decays and determine the corresponding
$CP$-averaged fit fractions and charge asymmetries.  
\vfill
\begin{center}

Submitted to the 33$^{\rm rd}$ International Conference on High-Energy Physics, ICHEP 08,\\
30 July---5 August 2008, Philadelphia, Pennsylvania.

\end{center}

\vspace{1.0cm}
\begin{center}
{\em Stanford Linear Accelerator Center, Stanford University, 
Stanford, CA 94309} \\ \vspace{0.1cm}\hrule\vspace{0.1cm}
Work supported in part by Department of Energy contract DE-AC02-76SF00515.
\end{center}

\newpage
} 

%
%
\begin{center}
\small

The \babar\ Collaboration,
\bigskip

%
B.~Aubert,
M.~Bona,
Y.~Karyotakis,
J.~P.~Lees,
V.~Poireau,
E.~Prencipe,
X.~Prudent,
V.~Tisserand
\inst{Laboratoire de Physique des Particules, IN2P3/CNRS et Universit\'e de Savoie, F-74941 Annecy-Le-Vieux, France }
J.~Garra~Tico,
E.~Grauges
\inst{Universitat de Barcelona, Facultat de Fisica, Departament ECM, E-08028 Barcelona, Spain }
% ...

\end{center}\newpage

\section{INTRODUCTION}
\label{sec:Introduction}
Amplitude analyses of three-body decays of \B mesons with no charm
particle in the final state are well suited to study the Cabibbo-Kobayashi-Maskawa (CKM)
framework~\cite{CKM} for charged current weak interactions. In the analysis of a Dalitz plot the strong phases between
interfering resonances are measured and can be used to constrain the 
weak phases related to the CKM parameters that, in the Standard Model, govern $CP$-violation.
Following the path~\cite{SnyderQuinn,rhopibabar,rhopibelle} of the 3-pion \B meson decays which give
constraints on the CKM angle $\alpha_{\rm CKM}\equiv \arg(-V_{td}V_{tb}^*/V_{ud}V_{ub}^*)$, it has been shown in~\cite{Ciuchini:2006kv,Gronau:2006qn} 
that \B decays into a kaon and two pions are sensitive to the angle $\gamma_{\rm CKM}\equiv \arg(-V_{ud}V_{ub}^*/V_{cd}V_{cb}^*)$. \par
In this paper we present an amplitude analysis of the flavor-specific $\Bz\to\Kp\pim\piz$ decay~\cite{Cmodeimplied}. This analysis, an update to an earlier analysis~\cite{josezhitang}, compares the Dalitz plots of the \Bz and \Bzb
decays where $K\pi$ and $\pi\pi$ resonances
interfere. In addition to enhanced statistics, we utilize improved track reconstruction, tagging information from the opposite B, and the measured B flight-time to improve the quality of the measurements. Previous measurements of the three-body final
state~\cite{belle,cleo} and subdecays~\cite{cleorhok,Feng} to a vector and a
pseudoscalar meson have been published. Other $\B\to K\pi\pi$ decays
have been studied in~\cite{latham, bellek+pi+pi-, babar-kspipi, belle-kspipi}. A phenomenological study of three-body \B meson
decays without charm in the final state is presented in~\cite{Cheng:2007si}. \par
This paper is organized as follows. We first present in~\secref{DecayAmplitudes} the decay model based on an isobar
expansion of the three-body decay amplitude. The complex coefficients of the expansion
are the unknowns we seek to determine by means of an unbinned extended maximum likelihood fit. 
We describe the detector and dataset in~\secref{DetectorAndData}, the procedure used to select the data sample in~\secref{EventSelection},
and the fit method in~\secref{TheFit}. The results are then described
in~\secref{Results} together with the accounting of the systematic
uncertainties in~\secref{Systematics}. Finally in~\secref{Summary}, we summarize our findings. 

\section{DECAY AMPLITUDES}
\label{sec:DecayAmplitudes}
The $\Bz\to\Kp\pim\piz$ decay amplitude is a function of two independent kinematic variables commonly chosen to be
the invariant masses squared\footnote{We use natural units where $\hbar=c=1$ in our algebraic equations}, 
$x=m_{\Kpm\pimp}^2$ and $y=m_{\Kpm\piz}^2$. The Dalitz plot (DP) is the $x,\ y$ two-dimensional distribution.  
It is customary to express the decay amplitude as a sum over intermediate (isobar) states:
\begin{equation}
\label{eq:isobarB}
\mathcal{A}(x,y)             = \sum_j a_j f_j(x,y), 
\end{equation}
and similarly for the $\Bzb\to\Km\pip\piz$ Dalitz plot,
\begin{equation}
\label{eq:isobarBbar}
\overline{\mathcal{A}}(x,y)  = \sum_j \overline{a}_j f_j(x,y).
\end{equation}
The complex isobar coefficients $a_j$ are parameterized by:
\begin{eqnarray}
\label{eq:isobarCoeff}
a_j = c_j(1+b_j)e^{i(\phi_j+\delta_j)}\\
\overline{a}_j = c_j(1-b_j)e^{i(\phi_j-\delta_j)}
\end{eqnarray}
and are constant over the Dalitz plot. The parameters $b_k,~\phi_k,~\delta_k$ are related to the isobar fractions $FF_k$ ($CP$-averaged over $\Bz$ and $\Bzb$), $CP$-violation charge asymmetries and phases by:
\small
\begin{eqnarray} \label{eq:PartialFractions}
  FF_k &=& \frac{\int_{DP}[|a_k f_k(x,y)|^2+|\overline{a}_k\overline{f}_k(x,y)|^2]dx\ dy} 
                {\int_{DP}[|\sum_ja_j f_j(x,y)|^2+|\sum_j\overline{a}_j\overline{f}_j(x,y)|^2]dx\ dy} \\
  A_{\rm CP}^k &=& \frac{|\overline{a}_k|^2 - |a_k|^2}
                         {|\overline{a}_k|^2 + |a_k|^2}~ =~	\frac{-2b_k}{1+b_k^2}~\nonumber\\
  \Phi_k & = & \phi_k+\delta_k~,\nonumber\\
  \overline{\Phi}_k & = & \phi_k-\delta_k \nonumber
\end{eqnarray}
\normalsize
\noindent Note that, due to interference, the fractions $FF_k$ in general do not add up to unity.

The decay dynamics of an intermediate state are specified by the $f_j(x,y)$ function which describes the Dalitz plot.
For instance a resonance formed in the $\Kp\pim$ system gives a contribution which factorizes as:
\begin{equation}
\label{eq:fequalRT}
f_j(x,y) = R_j(x) \times T_j(x,y) \times W_j(x),
\end{equation}
where $R_j(x)$ is the resonance mass distribution or lineshape and  $T_j(x,y)$ models the angular dependence. 
The product of Blatt-Weisskopf damping factors, $W_j(x)=\sqrt{B_B(Rp^*(x))\;B_j(Rq(x))}$~\cite{BlattWeissk}, slightly deviates from unity as
a function of $x$ through the breakup momenta\footnote{$p^*$, the momentum of the bachelor particle in the \B meson rest frame, is equal
to the breakup momentum of the studied \B meson decay.} of the (quasi) two body \B and resonance decays multiplied by a range parameter $R$.    
The $f_j$ are normalized,
\begin{equation}
\label{eq:isobarnorm}
\int_{DP}|f_j(x,y)|^2dx\ dy=1.
\end{equation}

\par
We use the Zemach tensor formalism~\cite{Asner,Zemach} for the angular distribution $T_j^{(J)}(x,y)$ of a process by which a pseudoscalar 
$\B$ meson produces a spin-$J$ resonance in association with a bachelor pseudoscalar meson. For $J=0,\ 1,\ 2$, we have:
\begin{eqnarray}
\label{zemach}
T_j^{(0)} &=& 1, \nonumber \\
T_j^{(1)} &=&  -2 \vec{p}\cdot\vec{q}, \nonumber \\
T_j^{(2)} &=&  \frac{4}{3} [3(\vec{p}\cdot\vec{q})^2 - (|\vec{p}||\vec{q}|)^2],
\end{eqnarray}
where\footnote{For simplicity, we have dropped the $j$ index in  $\vec{p}$ and $\vec{q}$.} $\vec{p}(x,y)$ ($\vec{q}(x)$) is the momentum vector of the bachelor particle 
(the resonance decay product $Q$ defined below) 
measured in the resonance rest frame. For a neutral (charged) $K\pi$ resonance, $Q$ is the pion (kaon), and for a dipion 
resonance, $Q$ is the $\piz$. Notice that these choices define for each two-body system the helicity angle  $\theta_j=(\vec{p_j},\ \vec{q_j})$ 
between 0 and $\pi$.\par
Our nominal model (\tabref{nominal}) for the decay $\Bz\to\Kp\pim\piz$ includes a nonresonant contribution which is uniformly distributed over the Dalitz plot, 
and seven resonant intermediate states: $\rho^-(770)\Kp$, $\rho^-(1450)\Kp$, $\rho^-(1700)\Kp$, $K^*(892)^{+,0}\pi^{-,0}$
and $(K\pi)^{*+,0}_0\pi^{-,0}$. The notation for the last isobar component, introduced by the \babar\ experiment~\cite{latham}, denotes phenomenological
amplitudes describing the neutral and charged $K\pi$ S-waves each by a coherent superposition of an elastic effective range term and a
term for the $\Kstar_0(1430)$ scalar resonance. It describes current knowledge on low energy $K\pi$ systems with a small number of parameters.  
In addition we include two non-interfering components for the decays $\Bz\to\overline{D}^0\piz\to\Kp\pim\piz$ and $\Bz\to D^-\Kp\to\Kp\pim\piz$.
Variations in the nominal model are used to estimate the model-dependent systematic uncertainty in the results. The Gounaris-Sakurai~(GS), relativistic Breit-Wigner~(RBW), and LASS lineshapes are used to model the $R_j(x)$. Parameters are taken from~\cite{PDG2006} unless stated otherwise. 

\begin{table}
\begin{center}
\caption{\label{tab:nominal}The {\it nominal model} for the decay $\Bz\to\Kp\pim\piz$ comprises a nonresonant part and seven intermediate states. The three types of 
lineshape are described in the text. The resonances masses and widths are from~\cite{PDG2006}, except for the LASS shape~\cite{LASS}. We use the same LASS 
parameters for both neutral and charged $K\pi$ systems. {\it Additional resonances} that may contribute are included in extended models which
we study to estimate the systematic uncertainties.}
\begin{tabular}{ccc} \hline\hline
Intermediate state & Lineshape & Parameters  \\ \hline
\multicolumn{3}{c}{{\it Nominal model}}\\ 
&& \\
Nonresonant & Constant & \\ 
&& \\
$\rho^-(770)$              & GS  & $m=768.5\ \ \ \mevcc$  \\                                                           
                           &     & $\Gamma^0=\ 148.2\ \ \ \mev\ \ $\\ 
$\rho^-(1450)$             & GS  & $m=1439\ \ \ \mevcc$ \\
                           &     & $\Gamma^0=\ 550\ \ \ \mev\ \ $ \\
$\rho^-(1700)$             & GS  & $m=1795\ \ \ \mevcc$ \\ 
                           &     & $\Gamma^0=\ 278\ \ \ \mev\ \ $ \\
&& \\
$\Kstarp(892)$             & RBW  &\\
$\Kstarz(892)$             & RBW  &\\ 
&& \\
$(K\pi)^{*+}_0$            & LASS & $m^0 = 1415 \pm 3\ \ \ \mevcc$ \\
$(K\pi)^{*0}_0$            &      & $\Gamma^0 = \ 300 \pm 6\ \ \ \mev\ \ $ \\
                           &      & cutoff $m_j^{max}=1800 \mevcc$  \\
                           &      & $a = 2.07 \pm 0.10 ~(\gevc)^{-1}$ \\ 
                           &      & $r = 3.32 \pm 0.34 ~(\gevc)^{-1}$\\ \hline
\multicolumn{3}{c}{{\it Non-interfering Components}} \\ 
$D^0$           &  $mass=1862.3$         & Double Gaussian    \\
                &  $width=7.1\mevcc$     & (From Data)              \\\hline
$D^+$           &  $mass=1864.4$         & Double Gaussian      \\
                &  $width=9.9\mevcc$     & (From MC)             \\\hline
\multicolumn{3}{c}{{\it Additional resonances}} \\ 
&& \\
$K_2^*(1430)^{+,0}$        & RBW &\\
$K^*(1680)^{+,0}$          & RBW & \\ \hline \hline
\end{tabular}
\end{center}
\end{table}

\subsection{LINESHAPES}
\label{subsec:lineshapes}
\subsubsection{The relativistic Breit-Wigner distribution}
The relativistic Breit-Wigner~(RBW) parameterization 
is used for $K^*(892)^{+,0}$, $K_2^*(1430)^{+,0}$, and $K^*(1680)^{+,0}$:
\beq
\label{eq:nominalBW}
        R^{(J)}_j(x;m_j,\Gamma_j^0) \;=\; 
                \frac{1}{m_j^2 - x - i m_j\Gamma^{(J)}_j(x)}~.
\eeq
The mass-dependence of the total width $\Gamma^{(J)}_j$ can be ignored for high-mass states. For the low-mass states which decay only elastically, it is defined by
\beq
\label{eq:s-dependentWidth}
        \Gamma^{(J)}_j(x) \;=\; 
                \Gamma_j^0
                \frac{m_j}{\sqrt{x}}
                \left(\frac{q(x)}{q(m_j^2)}\right)^{\!2J+1}
                \frac{B^{(J)}(Rq(x))}{B^{(J)}(Rq(m_j^2))}~,
\eeq
where $m_j$ is the mass of the resonance $j$, $\Gamma_j^0=\Gamma_j(m_j^2)$ 
its width, and the barrier factors (squares of the Blatt-Weisskopf damping factors~\cite{BlattWeissk}) are:
\beqn
\label{eq:barrier}
        B^{(0)}      &=& 1, \\
        B^{(1)}      &=& \frac{1}{1 + R^2 q^2}, \nonumber \\ 
        B^{(2)}      &=& \frac{1}{9 + 3 R^2 q^2 + R^4 q^4}. \nonumber
\eeqn
All range parameters~(R) are set to the values in the PDG~\cite{PDG2006}. 

\subsubsection{The Gounaris-Sakurai distribution}
The Gounaris-Sakurai~(GS) parameterization~\cite{rhoGS} 
is used for $\rho^-(770)$, $\rho^-(1450)$ and $\rho^-(1700)$:
\beq
\label{eq:GS}
R^{GS}_j(x;m_j,\Gamma_j^0) \;=\; \frac{1+d_j \;\Gamma^0_j/m_j}{m_j^2+g_j(x)-x -i m_j\Gamma_j(x)},
\eeq
with the same $x$-dependence of the width as for the RBW. The expressions of the constant $d_j$ and the function $g_j(x)$ in terms of 
$m_j$ and $\Gamma^0_j$ are given in~\cite{rhoGS}. The parameters of the $\rho$ lineshapes are taken from $\tau$ and $\pi\pi$ scattering in~\cite{CMD2pipi} and~\cite{rhoFitDM2}. 
\subsubsection{The LASS distribution}
For the $K\pi\, S$-wave amplitudes, $(K\pi)^{*+,0}_0$, which dominate for $m_{K\pi}$ below $m_j^{max}=2~\gevcc$, an effective-range 
parameterization was suggested~\cite{Estabrooks} to describe the slowly increasing phase as a function of the $K\pi$ mass.
We use the parameterization as in the LASS experiment~\cite{LASS}, tuned for \B decays:

\begin{eqnarray}
\label{eq:LASS}
R_j^{LASS} (x; m_j^0,\Gamma_j^0,a,r) =\frac{\sqrt{x}}{q \cot{\delta_B} - iq} \ \ \ \ \ \ \ \ \ \\ 
                    \ \ \ \ \ \ \ \ \ \ \ \ \ +e^{2i \delta_B} \frac{m_j^0 \Gamma_j^0 \frac{m_j^0}{q_0}}{[(m_j^0)^2 - x] - i m_j^0 \Gamma_j^0 \frac{q}{\sqrt{x}} \frac{m_j^0}{q_0}},\nonumber
\end{eqnarray}
\noindent where
         \begin{equation}
          \label{eq:LASSphase}
           \cot{\delta_B} = \frac{1}{a q(x)} + \frac{1}{2}\,r\,q(x)\,,
         \end{equation}
        $a$ is the scattering length, and $r$ the effective range (\tabref{nominal}).

\subsection{THE SQUARE DALITZ PLOT}

\begin{figure}
  \epsfig{file=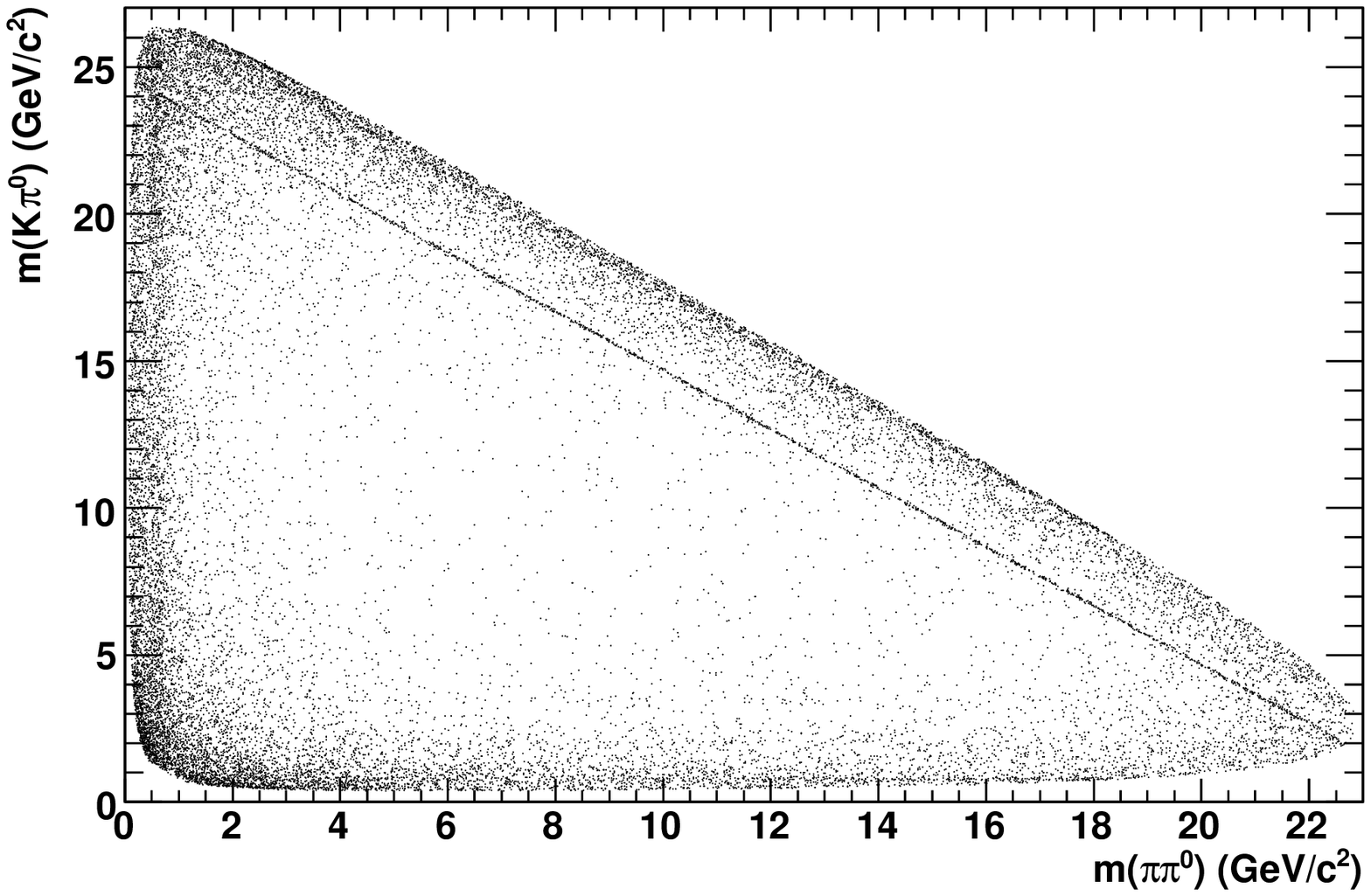,width=7cm}
  \epsfig{file=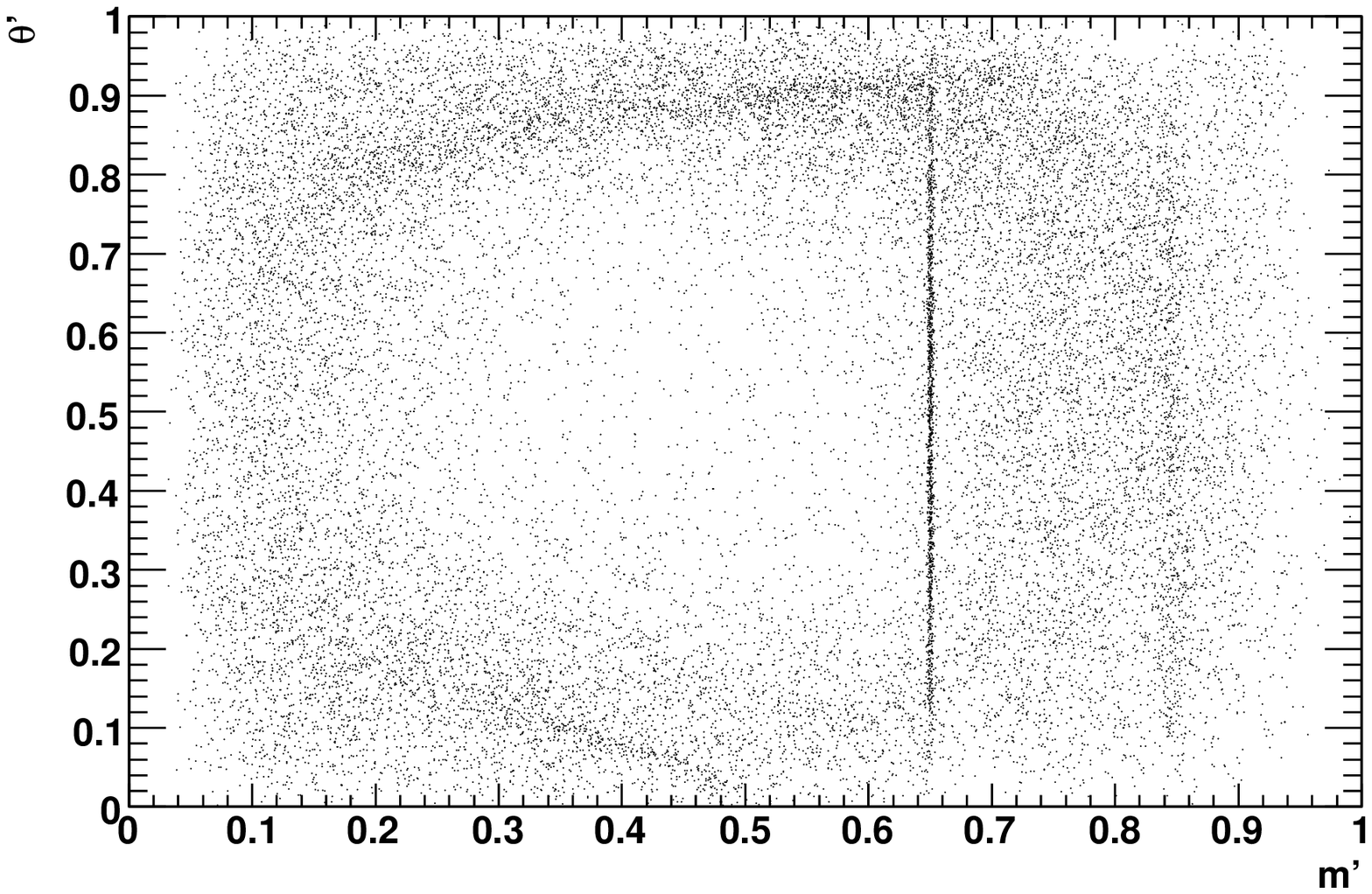,width=7cm}
  \caption {\label{fig:DalitzPlots}
    The standard (a) and square (b) Dalitz plots of the
    selected data sample of 23683 events. The selection criteria are described in Sec.~\ref{sec:EventSelection} . The structures
    are more spread out in the square Dalitz plot. The
    $\Dzb\to\Kp\pim$ narrow band is preserved with the
    choice made for the \mprime\ variable.}
\end{figure}

The accessible phase space for charmless three-body \B decays is unusually large. Most contributing resonances have masses much lower than the \B mass. 
Hence signal events cluster along the Dalitz plot boundaries. This is also true for background events. 
Past experience has shown that another set of variables, defining the {\it Square Dalitz Plot} (SDP) is well suited to such configurations. 
It is defined by the mapping:

\begin{eqnarray}
\label{eq:SDPvariables}
dx\ dy &\longrightarrow &d\mprime\ d\thetaprime \\
\ \ \mprime&\equiv&\frac{1}{\pi}\arccos(2\frac{m-m_{\min}}{m_{\max}-m_{\min}}-1)\;,\;\thetaprime\equiv\frac{1}{\pi}\theta, \nonumber
\end{eqnarray}
where $m=\sqrt{x}$ and $\theta$ are respectively the invariant mass and helicity angle of the $K^{\pm}\pi^{\mp}$ system. 
$m_{\max}=m_B-m_{\piz}$ and $m_{\min}=m_{\Kp}+m_{\pim}$ are the kinematic limits of $m$. The new variables both range between 0 and 1. 
The standard and square Dalitz plots are shown for our data sample in~\figref{DalitzPlots}.

\section{THE \babar\ DETECTOR AND DATASET}
\label{sec:DetectorAndData}

The data used in this analysis were collected with the \babar\ detector
at the \pep2\ asymmetric energy \epem\ storage rings between October 1999 and September 2007.
This corresponds to an integrated luminosity of 413 \invfb\ or approximately $N_{\BB}=454\pm5$ million $\BB$ pairs taken on the peak of the $\FourS$ resonance (on resonance) and 41\invfb\ recorded at a center-of-mass (CM) energy 40~$\mev$ below (off resonance).\par
A detailed description of the \babar\ detector is given in~\cite{babarNim}. 
Charged-particle trajectories are measured by a five-layer, double-sided silicon vertex tracker (SVT) and a 40-layer
drift chamber (DCH) coaxial with a 1.5~T magnetic field. Charged-particle identification is achieved by combining
the information from a ring-imaging Cherenkov device (DIRC) 
with the ionization energy loss (\dedx ) measurements 
from the DCH and SVT. Photons are detected in a CsI(Tl) electromagnetic calorimeter (EMC) inside the coil. 
Muon candidates are identified in the instrumented flux return 
of the solenoid.  We use GEANT4-based~\cite{geant4} software to simulate the detector response and account for the varying beam and environmental conditions. Using this software, we generate signal and background Monte Carlo (MC) to estimate the efficiency and expected backgrounds in this analysis.  Two samples of signal MC were used:  one was generated with the Dalitz plot distribution observed in the previous analysis~\cite{josezhitang} while the other was generated with a phase-space distribution.

 \section{EVENT SELECTION}
\label{sec:EventSelection}
\subsection{SIGNAL SELECTION AND BACKGROUND REJECTION}
\label{subsec:Cuts}

To reconstruct $\Bz\to\Kp\pim\piz$ decays, we select two charged particles and two photons. The charged particle candidates are required to have 
transverse momenta above $100\ \mevc$ and at least 12 hits in the DCH. They must not be identified as electrons or muons or protons. 
We select kaons and pions based on their signatures in the DIRC and DCH. The $\piz$ candidate is built from a pair of photon candidates, each with an energy greater
than $50\mev$ in the laboratory frame (LAB) and a lateral energy deposition profile in the EMC consistent with an electromagnetic shower. 
The invariant mass of a $\piz$ candidate must satisfy $\vert \frac{m_{\pi^0} -  m_{\rm PDG}}{\sigma_{m_{\pi^0}}} \vert < 3$. We also require $|\cos\theta^{*}_{\pi^0}|$, the modulus of the cosine of the angle the decay photons make with the $\pi^0$ momentum vector to be less than 0.95 .\par 
At the $\FourS$ resonance, $\B$ mesons are characterized by two nearly independent kinematic variables, the beam energy substituted mass and the energy 
difference:
\beqn
\label{eq:mesde}
\mes&=&\sqrt{(s/2+\vec{p_0} \cdot \vec{p_B})^2/E_0^2-p_B^2}, \\
\Delta E&=&E_B^*-\sqrt{s}/2,
\eeqn
where $E$ and $p$ are energy and momentum, the
subscripts 0 and $B$ refer to the \epem-beam system and the \B candidate respectively; $s$ is the square of the center-of-mass energy and the asterisk 
labels the CM frame. We require that $5.272<\mes<5.2875\gevcc$. To avoid a bias in the Dalitz plot from the dependence on the $\piz$ energy of the
resolution in \de,  we introduce the dimensionless quantity: 

\beqn
\label{eq:deprime}
    \deprime = \frac {\frac{\de} {\sigma_{\de}} + m_0 + m_1x + m_2x^2 + m_3x^3}  {
w_0 + w_1x + w_2x^2 + w_3x^3}
\eeqn
 
\noindent where the  coefficients are determined from fits to signal MC and $x=m_{\Kpm\pimp}^2$. We require $|\deprime| \leq 2.1$.

Continuum $e^+e^-\to q\bar{q}$ ($q = u,d,s,c$) events are the dominant  background. To enhance discrimination between signal and continuum, 
we select events by using a neural network~\cite{NN} with an output $NN$ which combines six discriminating variables: the angles of the \B momentum and the \B thrust axis 
with respect to the $e^+$ beam direction in the CM frame, the angle between the thrust axes of the signal \B and other \B, the  zeroth and second order monomials $L_0$ and $L_2$, and $\Delta z/\sigma(\Delta z)$, the flight distance between the two {\B}s scaled by the error. The monomials are defined as $L_n\equiv\sum_i p_i\cdot|\cos\theta_i|^n$, where the sum runs over all charged and neutral particles in the event 
(except for those in the \B candidate) whose momenta $\vec{p_i}$ make angles $\theta_i$ with the $\B$ thrust axis. 
The neural network was trained on {\it off  resonance} data and correctly reconstructed signal
Monte Carlo events. We require $0.6 < NN$. \par

Approximately 15\% of the signal events have multiple reconstructed $\B$ candidates (usually two).
We select the candidate with the minimum value of: 
  \beqn
  \chi^2 = \bigl( \frac{m_{\pi^0} -  m_{\rm PDG}}{\sigma_{m_{\pi^0}}} \bigr)^2 + \chi_{\rm Vertex}^2.  
  \eeqn
where  $\chi_{\rm Vertex}^2$ is the $\chi^2$ of the kinematic fit to the particles in the \B\ meson candidate.  

There are 23268 events in the data sample after the selection.  The \B\ meson candidate in each event is mass constrained to ensure that the measurement
falls within the Dalitz plot boundary. 

\subsection{TRUTH-MATCHED AND SELF-CROSS-FEED SIGNAL EVENTS}
\begin{figure}[t]
  \centerline{\epsfxsize7cm\epsffile{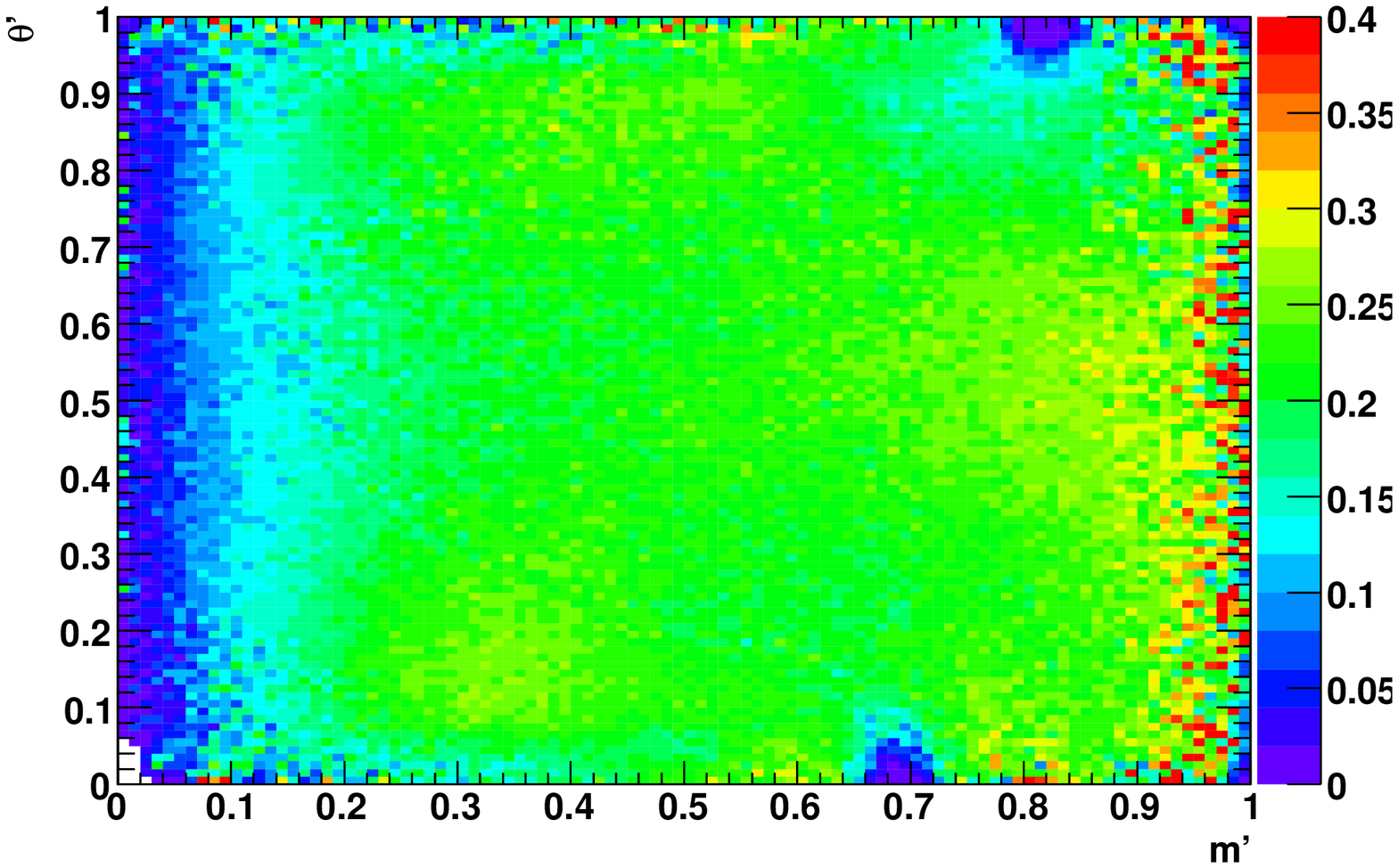}
              \epsfxsize7cm\epsffile{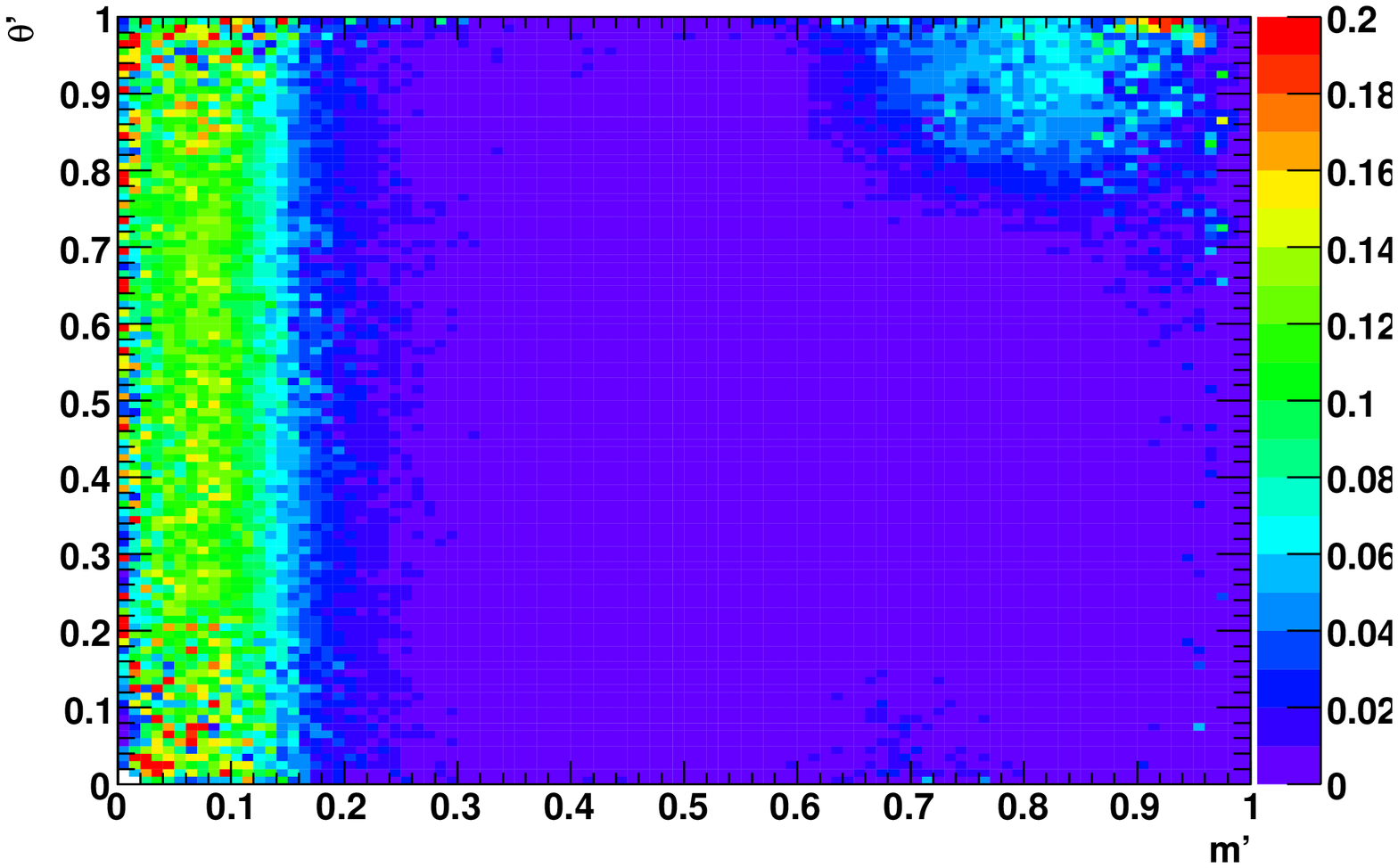} }
  \vspace{-0.4cm}
 \caption{\label{fig:efficiency}
           \em Selection efficiency for truth-matched events on the left and
               SCF events on the right.}
\end{figure}

Using the Monte Carlo simulation as in~\cite{rhopibabar}, we distinguish between the correctly reconstructed and the misreconstructed signal events.  
A correctly reconstructed event where the three particles of the \B candidate match the generated ones, is called a {\it Truth-Matched} (TM) event. The TM PDFs describe correctly reconstructed events in the fit to data. A misreconstructed signal event contains a \B meson  which decays to the signal mode, but one or more reconstructed particles in the \B candidate are not actually from the decay of that \B. Misreconstructed signal is called  {\it Self-Cross-Feed} (SCF). 
Misreconstruction is primarily due to the presence of low momentum pions. Consequently the efficiency $\varepsilon(\mprime,\thetaprime)$ to reconstruct an event 
either correctly or incorrectly varies across the Dalitz plot. The SCF fraction~$f_{\rm SCF}(\mprime,\thetaprime)$ is high, where the quality of the reconstruction is poor. This occurs in the corners of the Dalitz plot where one of the final-state particles has a low momentum in the LAB frame. These variations can be seen 
in Fig.~\ref{fig:efficiency} computed using high statistics Monte Carlo samples. It is important to keep a high efficiency in the Dalitz plot corners where the low-mass vector resonances interfere. Overall the total efficiency is close to 22.5\% and the SCF fraction, averaged over the Dalitz plot, is $\sim 9\%$.

\label{subsec:tmscf}

\subsection{BACKGROUND}
\subsubsection{Continuum background}
\label{subsubsec:continuum}
Although the neural network selection rejects 90\% of the continuum events, this background is the dominant class of events in the data sample, representing about two thirds of its size. 
\subsubsection{Background from other \B decays}
\label{subsubsec:bbackground}
Since there is no restriction on any two-body invariant mass of the
final state particles, large backgrounds from other \B decays
occur. We use high statistics Monte Carlo samples to study these
backgrounds. Conservative assumptions about unknown branching
fractions are made. Inclusive and exclusive \B decays with or without
charm are grouped into nineteen classes to be used in the fit. Rates, and
topological and kinematical similarities are studied to define  
the classes listed in~\tabref{listBb}. Those backgrounds whose contributions are expected to 
be large (200 or more events) are varied in the fit while all others are fixed. 

\begin{table}[h]
\begin{center}
\caption{\label{tab:listBb} The list of \B-backgrounds retained for the fit (\secref{TheFit}). For each channel, we
give (anticipating~\secref{Results}) either the fitted number of events in the data sample if 
its yield is allowed to vary in the fit procedure or the expected number otherwise.}

\begin{tabular}{clrr}\hline\hline
Class   & Mode  &Events&  \\ \hline
1	& $\Bz\to\Kstarz(892)\g,\Kstarz(1430)\g$                     &$187 \pm 14$  & fixed   \\
2	&  $\rho^+\piz$                                             &$11\pm 2$  & fixed   \\
3	&  $f_0(980)K^+$, $K^{*+}\piz$,                             &$48\pm 12$  & fixed   \\
4	& $K^{*+}K^-$                                               &$14\pm 10$  & fixed   \\
5	& $\pip\pim\pip$ Dalitz plot                                &$8\pm 1$  & fixed   \\
6	&   $K^+\pim\pip$ Dalitz plot   			    &$164\pm 9$  & fixed   \\
7	& $K^+\piz$						    &$65 \pm 3$  & fixed   \\ 
8	& $K^+\pim$						    &$53 \pm 2$  & fixed   \\ 
9	& $\pip\pim\piz$ Dalitz plot                                &$109\pm 13$  & fixed   \\
10	& Generic $\B\to$ charm with  $\Dz$                         &$627\pm60$ & varied \\  
11	& Generic $\B\to$ charm with  $\Dp$                         &$370\pm80$ & varied \\  
12	&  $K^{*+}a_1^-$, $K^{*0}\rho^0$,                           &$9\pm 2$  & fixed   \\
13	&  $K^+\eta\pi^-$,                            		    &$8\pm 1$  & fixed   \\
14	&  $\eta^\prime K^+$,                                       &$22\pm 1$  & fixed   \\
15	&  $\rho^+\rho^-$,$ a_1^+\pi^-$                 	    &$27\pm 3$  & fixed   \\
16	&  $K^{*0}\rho^+$,                                          &$15\pm 6$  & fixed   \\
17	&  $K^{*+}\rho^-$,                                          &$21\pm 6$  & fixed   \\
18	&  $\rho^+\rho^0$,$a_1^0\pip$,$a_1^+\piz$                  &$50\pm 13$  & fixed   \\
19      &  Combinatoric \B Decays  			            &$660\pm122$&varied  \\
\hline\hline
\end{tabular}
\end{center}
\end{table}
 
\section{THE MAXIMUM LIKELIHOOD FIT}
\label{sec:TheFit}
We perform an unbinned extended maximum likelihood fit to determine the total $\Bz\to\Kp\pim\piz$ event yield, the magnitudes 
$c_j(1\pm b_j)$ and phases $\phi_j \pm \delta_j$ of the complex isobar coefficients of the decay amplitude defined in Eq.~\ref{eq:isobarCoeff}.
The fit uses the variables \mprime, \thetaprime, \mes, \deprime\, and  NN to discriminate signal from background. A simultaneous fit is performed using the B-tagging~\cite{btoccbarK} category from the opposite B, for a further improvement in discriminating power. The variable label $\cat$ denotes each of seven tagging categories defined in~\cite{btoccbarK}.   

\subsection{THE LIKELIHOOD FUNCTION}
\label{subsec:thelikeli}
The selected on-resonance data sample consists of signal, continuum-background and background from other \B decays. 
The probability density function (PDF) ${\cal P}_i^\cat$ for an
event $i$ in tagging category $\cat$ is the sum of the probability densities 
of all components, namely

\begin{eqnarray}
\label{eq:theLikelihood}
        {\cal P}_i^\cat
        &\equiv& 
                N_{\rm sig} f^\cat_{\rm sig}
                \left[  (1-\fscfave^\cat){\cal P}_{{\rm sig}-\TM,i}^\cat +
                        \fscfave^\cat{\cal P}_{{\rm sig}-\SCF,i}^\cat 
                \right] 
                \nonumber\\[0.3cm]
        &&
                +\; N^\cat_{q\bar q}\frac{1}{2}
                \left(1 + \Qtagi\Atagqq\right){\cal P}_{q\bar q,i}^\cat
                \nonumber \\[0.3cm]
        &&
                +\; \sum_{j=1}^{N^{B,j}_{\rm class}}
                N_{B,j} f^\cat_{B,j}
                \frac{1}{2}\left(1 + \Qtagi \Atagj\right){\cal P}_{B,ij}^\cat
                \nonumber \\[0.3cm]
\end{eqnarray}
where
        $N_{\rm sig}$ is the total number of $\Bz\to\Kp\pim\piz$ signal events 
        in the data sample;
        $f^\cat_{\rm sig}$ is the fraction of signal events that are 
        tagged in category $\cat$;
        $\fscfave^\cat$ is the fraction of SCF events in tagging category $\cat$, 
        averaged over the DP;
        ${\cal P}_{{\rm sig}-\TM,i}^\cat$ and ${\cal P}_{{\rm sig}-\SCF,i}^\cat$
        are the products of PDFs of the discriminating variables used
        in tagging category $\cat$ for TM and SCF
        events, respectively; 
        $N^\cat_{q\bar q}$ is the number of continuum events that are 
        tagged in category $\cat$;
        $\Qtagi$ is the tag flavor of the event, and is equal to the 
	charge of the kaon from the $B$ decay; 
        $\Atagqq$ parameterizes possible tag asymmetry in continuum events; 
        ${\cal P}_{q\bar q,i}^\cat$ is the continuum PDF for tagging 
        category $\cat$;
        $N^{B}_{\rm class}$ is the number of 
         $B$-related background classes considered in the fit,
        namely nineteen;
        $N_{B,j}$  is the number of expected events in
        the  $B$ background class $j$;
        $f^\cat_{B,j}$  is the fraction of 
         $B$ background events of class $j$
        that are tagged in category $\cat$;
        $\Atagj$ describes a possible tag asymmetry in the $B$ background
        class $j$;     
        ${\cal P}_{B,ij}^\cat$ is the $B$-background PDF for tagging 
        category $\cat$ and class $j$.

The PDFs ${\cal P}_{X}^{\cat}$ ($X=\{{\rm sig}\!-\!\TM,\  {\rm sig}\!-\!\SCF,\  q\bar q,\  B$)
are the product of the four PDFs of the discriminating variables~\footnote{
Not all the PDFs depend on the tagging category.
The general notations $P_{X,i(j)}^\cat$ and ${\cal P}_{X,i(j)}^{\cat}$ are used for simplicity.},
$x_1 = \mes$, $x_2 = \deprime$, $x_3 = {\rm NN~output}$ and the doublet
$x_4 = \{\mprime, \thetaprime\}$:
\begin{equation}
\label{eq:likVars}
        {\cal P}_{X,i(j)}^{\cat} \;\equiv\; 
        \prod_{k=1}^4 P_{X,i(j)}^\cat(x_k)~,
\end{equation}
where $i$ is the event index and $j$ is a $B$ background class.  The extended likelihood
over all tagging categories is given by
\begin{equation}
        {\cal L} \;\equiv\;  
        \prod_{\cat=1}^{7} e^{-\overline N^\cat}\,
        \prod_{i}^{\overline N^\cat} {\cal P}_{i}^\cat~,
\end{equation}
where $\overline N^\cat$ is the total number of events expected in category 
$\cat$. 

The correlations among the measurements are handled by building conditional PDFs where appropriate. The PDF parametizations are given in~\tabref{PDF}, and a summary of the parameters varied in the fit can be found in Section V.D.
\begin{table*}
\begin{center}
\caption{\label{tab:PDF} Summary of the PDF parameterizations. G=Gaussian, P1=1st order polynomial, NP=non-parametric. 
The notation (DP) designates a PDF with parameters which vary over the Dalitz plot. The Dalitz plot signal model is described 
in~\secref{DecayAmplitudes}.}
\setlength{\tabcolsep}{1.2pc}
\begin{tabular}{ccccc}\hline\hline
Component      & $\mes$   & $\deprime$   & NN       & Dalitz \\ \hline
signal (TM)    &see text  & G(DP)+P1(DP) & NP       & see text \\
signal (SCF)   &NP        & NP           & NP       & see text \\
Continuum      &Argus     & P1           & see text & NP  \\ 
\B backgrounds &NP        & NP           & NP       & NP       \\
\hline\hline
\end{tabular}
\end{center}
\end{table*}  
\subsection{THE DALITZ PROBABILITY DENSITY FUNCTIONS}
\label{subsec:DalitzPDF}
Since the decay $\Bz\to\Kp\pim\piz$ is flavor-specific (the charge of the kaon identifies the $b$ flavor), the \Bz and \Bzb Dalitz plots are independent.
However, because the backgrounds are essentially flavor blind, we get a more robust procedure by fitting them simultaneously. 
It is enough to describe only the \Bz\ Dalitz plot PDF. A change from $\mathcal{A}$ to $\mathcal{\overline{A}}$ 
(\equtworef{isobarB}{isobarBbar}) accompanied by the interchange of the charges of the kaon and pion gives the \Bzb\ PDF.
\subsubsection{Signal}
The model for the distribution of signal events in the Dalitz plot has been described in \secref{DecayAmplitudes}. The free parameters are $c_j,~b_j~,\phi_j,~\delta_j$ defined in~\equtworef{isobarB}{isobarBbar} for all the intermediate states of the signal model given in~\tabref{nominal}. 
Since the measurement is done relative to the $\rho^-(770)$ final state, the phases of this and the charge conjugate channels are fixed to zero. The amplitude of $\Bz\to\rho^-(770)K^+$ is also fixed but not that of $\Bzb\to\rho^+(770)K^-$ in order to be sensitive to direct $CP$-violation. The weak phase $\delta_j$ and CP violating amplitude $b_j$ of the $\rho^-(1450)$ and $\rho^-(1700)$ are constrained to equal those of the $\rho^-(770)$ in the fit. \par

The normalization of the component signal PDFs:
\beq
\label{eq:unnormalizedTMDalitzPDFs}
\mathcal{P}_{TM,i}  \propto \varepsilon_i (1-f_{\SCF,i})|\mathrm{\mathrm{det}}\mathcal{J}_i||\mathcal{A}_i|^2,
\eeq
\beq
\label{eq:unnormalizedSCFDalitzPDFs}
\mathcal{P}_{\SCF,i} \propto \varepsilon_i   f_{\SCF,i}  [|\mathrm{det}\mathcal{J}||\mathcal{A}|^2 \otimes R_{\SCF}]_i,
\eeq
is model dependent. $\mathcal{J}$ is the Jacobian matrix of the mapping to the square Dalitz plot. 
The symbol $\otimes$ stands for a convolution and the $R$ matrix is described below in~\equaref{theMatrix}. 
The normalization requires the computation of the integrals 
\beqn
\label{eq:partialDPnorm}
\int_0^1d\mprime\int_0^1d\thetaprime\  \varepsilon (1-f_{\SCF})|\mathrm{det}\mathcal{J}|f_kf_l^*,\\
\int_0^1d\mprime\int_0^1d\thetaprime\  \varepsilon f_{\SCF}|\mathrm{det}\mathcal{J}|f_kf_l^*,
\eeqn
and
\beq
\label{eq:fullDPnorm}
\int_0^1d\mprime\int_0^1d\thetaprime\  \varepsilon|\mathrm{det}\mathcal{J}|f_kf_l^*,
\eeq
where the notations of ~\equaref{isobarB} are used. 
The integrations over the square Dalitz plot are performed numerically. 
The weight
\beq
\label{eq:fscfbar}
\overline{f}_{\SCF}= 
\frac{\int_0^1d\mprime\int_0^1d\thetaprime\  \varepsilon f_{\SCF}|\mathrm{det}\mathcal{J}||\mathcal{A}|^2}
     {\int_0^1d\mprime\int_0^1d\thetaprime\  \varepsilon|\mathrm{det}\mathcal{J}||\mathcal{A}|^2}
\eeq
ensures that the total signal PDF is normalized. The PDF normalization depends on the decay dynamics and is computed iteratively.
In practice the computation of $\overline{f}_{SCF}$ rapidly converges to a value which we fix after a few exploratory fits.  \par
Studies in simulation have shown that the experimental resolutions of \mprime\ and \thetaprime\ need not be introduced in the TM signal PDF. However, 
misreconstructed events often incur large migrations, when the reconstructed $\mprime_r, \thetaprime_r$ are far from the true values $\mprime_t, \thetaprime_t$. 
We use the Monte Carlo simulation to compute a normalized two-dimensional resolution function
$R_{SCF}(\mprime_r,\thetaprime_r;\mprime_t,\thetaprime_t)$, with
\beq
\label{eq:theMatrix}
\int_0^1d\mprime_r\int_0^1d\thetaprime_r R_{\SCF}(\mprime_r,\thetaprime_r;\mprime_t,\thetaprime_t)=1. 
\eeq 
$R_{SCF}$ is convolved with the signal model in the expression of $\mathcal{P}_{SCF}$ in~\equaref{unnormalizedSCFDalitzPDFs}.
\subsubsection{Background}
Except for events coming from exclusive $\B\to D$ decays, all background Dalitz PDF are modeled with non-parametric, smoothed, two-dimensional histograms. 
The continuum distributions are extracted from a combination of off resonance data and a sideband ($5.20<\mes<5.25\gevcc$) of the on-resonance 
data from which the \B-background has been subtracted. 
The square Dalitz plot is divided into eight regions where different smoothing parameters are applied in order to optimally reproduce the observed wide and 
narrow structures by using a two-dimensional kernel estimation technique~\cite{cranmer}. 
For $0.64 < \mprime < 0.66$ and all $\thetaprime$, a finely binned, unsmoothed histogram is used to follow the peak from the narrow $\Dz$ continuum production. 
The \B-background (\tabref{listBb}) Dalitz PDFs are obtained from the Monte Carlo simulation. For the components which model 
$b\to c$ decays with real $\Dz$ mesons, a fine grained binning around the $D$ mass is used to construct unsmoothed histograms.\par

\subsection{THE OTHER PDFS}
\label{subsec:nondalitzPDFs}
\subsubsection{Signal}

The $\mes$ distribution for signal events is parameterized as:

\begin{equation}
  \label{eqn:cruijff}
  f(x=\mes)=\exp\left[ -\frac{(x-m)^2} {2\sigma^2_\pm+\alpha_\pm(x-m)^2}\right]
\end{equation}

\noindent where $m$ and $\sigma_\pm$ are floated in the data fit.

For SCF-signal events we use a non-parametric shape taken from the Monte Carlo simulation. \par
\deprime\ is correlated with the Dalitz plot variables for TM-signal events. To account for the correlation, we choose the combination of a Gaussian and 1st order polynomial PDF. The mean and standard deviation of the Gaussian and slope of the polynomial vary linearly with $m_{\Kpm\pimp}^2$. These parameters (intercept and slope) are free in the fit. A non-parametric shape taken from the Monte Carlo simulation is used for the 
SCF-signal \deprime\ PDF. \par
The NN PDFs for TM and SCF events are non-parametric distributions taken from the Monte Carlo. 
\subsubsection{Background}
We use the Argus function~\cite{Argus}
\beq
\label{eq:argus}
f(z=\frac{\mes}{m_{\rm ES}^{\rm max}})\propto z\sqrt{1-z^2} e^{-\xi(1-z^2)}
\eeq 
as the continuum \mes\ PDF. The endpoint $m_{\rm ES}^{\rm max}$ is fixed to $5.2897\gevcc$ and $\xi$ is free in the fit. 
The \deprime\ PDF is a linear polynomial whose slope is free to vary in the fit. The shape of the NN distribution for 
continuum is correlated with the event location in the Dalitz plot. To account for that effect we use for the NN PDF a function that varies with the closest distance $\Delta_{\mathrm{dalitz}}$ between the point representing the event and the boundary of the standard Dalitz plot,
\beqn
    \label{eq:NNDP}
    \mathcal {P} (NN ;\Delta_{\mathrm{dalitz}}) = &&(1-NN)^{k_1} \\
                                                  &&\times (k_2 NN^2 + k_3 NN + k_4). \nonumber \\
                                                  && k_i = q_i + p_i\cdot \Delta_{\mathrm{dalitz}} \nonumber
\eeqn
The $k_i$ are linear functions of $\Delta_{\mathrm{dalitz}}$ where the $q_i$ and $p_i$ are varied in the likelihood fit. 

We use non-parametric distributions taken from the Monte Carlo to describe \mes, \deprime and NN distributions for the \B-background classes in~\tabref{listBb}. 
\subsection{THE FIT PARAMETERS}
\label{subsec:fitparam}
The following parameters are varied in the fit:
\begin{itemize}
\item Yields for signal ($N_{sig})$, continuum ($N_{\qqbar}$) and  three \B\ background classes (c=10, 11 and 19 defined in~\tabref{listBb}).
\item $CP$-asymmetries for the continuum events.
\item The global mean and slope(s), of the \deprime\ distribution for the TM-signal (continuum) events.  
\item Parameters which describe the shape and correlation of the NN output and the event location in the Dalitz plot [\equaref{NNDP}].
\item The mean and widths of the function describing the \mes distribution of the TM-signal events in addition to the $\xi$ parameter of the Argus function describing the continuum \mes shape. 
\item Thirty-two isobar magnitudes and phases. There are 10 intermediate states (7 resonances and a nonresonant term and two non-interfering D modes) and two Dalitz plots. We fix one reference magnitude, that of $\Bz\to\rho^-(770)K^{+}$ and two phases for the latter and its conjugate. 
      Therefore we end up with 18 magnitudes and 14 phases to be determined by the fit. 
\end{itemize}
\section{RESULTS}
\label{sec:Results}

The maximum likelihood fit results in a $\Bz \to \Kp\pim\piz$ event yield of $N_{sig}=4583\pm122$ events, where the uncertainty is statistical only. 
When the fit is repeated starting from input parameter values randomly chosen within wide ranges of one order of magnitude above and below the nominal values for the amplitudes and within the [$-\pi, \pi$] interval for the phases, we observe convergence toward four
solutions with minimum values of the negative loglikelihood function (NLL). The best solution is 
separated by 3.9 units of NLL from the next best solution. The event yield we quote is for the best solution; the spread of signal yields between the four solutions is less than 5 events. The fitted phases $\Phi$, $\overline{\Phi}$ and the $CP$-asymmetries $A_{CP}$ are given for the best solution in~\tabref{bestsolution}.\par

\begin{table}[h]
\caption{\label{tab:bestsolution}
Results of the best solution. The fractions are the $CP$-averaged isobar fractions ($FF_k$) defined with the $CP$-asymmetries $A_{CP}$ in~\secref{DecayAmplitudes}~[\equaref{PartialFractions}]. The phases $\Phi$ for the $\Bz$ decays and $\overline{\Phi}$ for the $\Bzb$ decays are measured relative to $\Bz(\Bzb)\to\rho^\mp K^\pm$. The first error is statistical and the second is systematic. }
\setlength{\tabcolsep}{1.2pc}
\begin{tabular}{ccccc}
\hline\hline
                        & Isobar Fraction ($\%$)                      &  $\overline{\Phi}$                              &  $\Phi$                                & $A_{CP}$           \\ 
\hline \\
$\rho^{-}(770)K^{+}$ 	& $ 13.60 \pm 1.24  \pm 0.60  $   & $ 0.00~(\mathrm{fixed})  $  	                   &  $ 0.00~(\mathrm{fixed})  $                          & $ 0.14 \pm  0.06  \pm 0.01 $     \\
$\rho^{-}(1450)K^{+}$ 	& $ 4.66  \pm 1.42  \pm 0.68  $   & $ 1.63 \pm  0.26  \pm 0.12  $&  $1.63  \pm  0.26  \pm 0.12 $  & $ 0.14 \pm  0.06  \pm 0.01 $   \\
$\rho^{-}(1700)K^{+}$ 	& $ 1.16  \pm 0.69  \pm 0.26  $   & $ 0.52 \pm  0.40  \pm 0.15  $&  $0.52  \pm  0.40  \pm 0.15 $  & $ 0.14 \pm  0.06  \pm 0.01 $   \\
$K^{*+}(892)\pi^{-}$ 	& $ 5.52  \pm 0.59  \pm 0.18  $   & $ 0.74 \pm  0.36  \pm 0.14  $&  $0.37  \pm  0.36  \pm 0.31 $  & $ -0.30  \pm  0.11  \pm 0.03 $    \\
$K^{*0}(892)\pi^{0}$ 	& $ 4.53  \pm 0.57  \pm 0.26  $   & $ 0.65 \pm  0.29  \pm 0.10  $&  $-0.00 \pm  0.33  \pm 0.10 $  & $ -0.15  \pm  0.12  \pm 0.02 $  	  \\
$(K\pi)^{*+}_0\pi^{-}$ 	& $ 23.60 \pm 1.18  \pm 1.70  $   & $-2.76 \pm  0.25  \pm 0.08  $&  $-2.60 \pm  0.30  \pm 0.22 $  & $ 0.07 \pm  0.05  \pm 0.01 $           \\
$(K\pi)^{*0}_0\pi^{0}$ 	& $ 11.90 \pm 1.11  \pm 1.46  $   & $ 0.37 \pm  0.26  \pm 0.38  $&  $-0.11 \pm  0.27  \pm 0.24 $  & $ -0.16  \pm  0.09  \pm 0.04 $          \\
N.R. 			& $  5.90 \pm 0.93  \pm 0.80  $   & $ 1.00 \pm  0.24  \pm 0.17  $&  $1.15  \pm  0.27  \pm 0.18 $  & $ 0.07 \pm  0.15  \pm 0.04 $          \\
$\overline{D}^0\pi^0$ 	& $ 20.90 \pm 0.85  \pm 2.66  $   &                                           &                                             &  	  \\
$D^- K^+$ 		& $  0.93 \pm 0.23  \pm 0.02  $   &                                           &                                             & 	  \\
\hline \\
\end{tabular}
\end{table}

\begin{figure}[h]
\epsfig{file=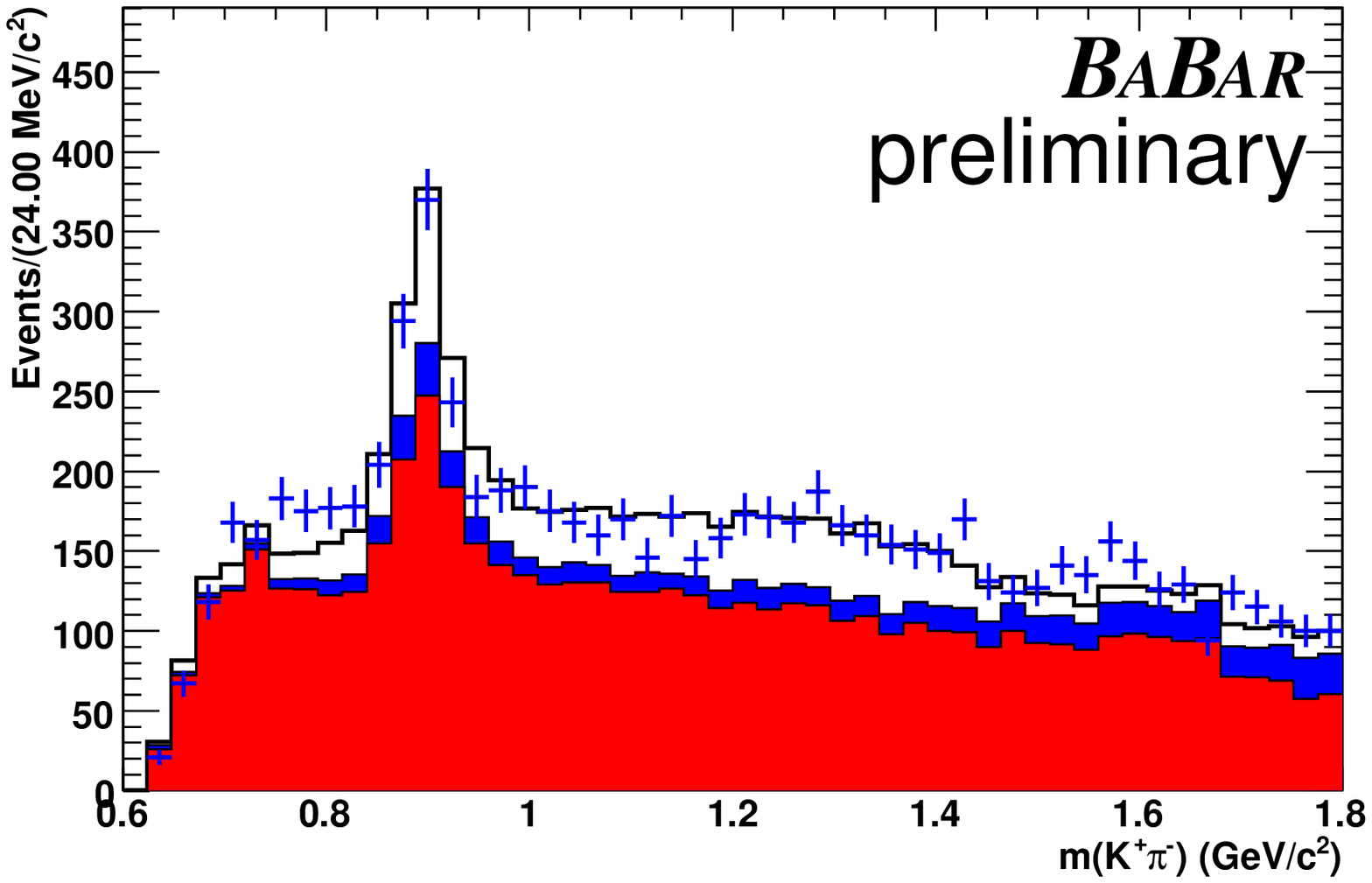,width=7cm}  
\epsfig{file=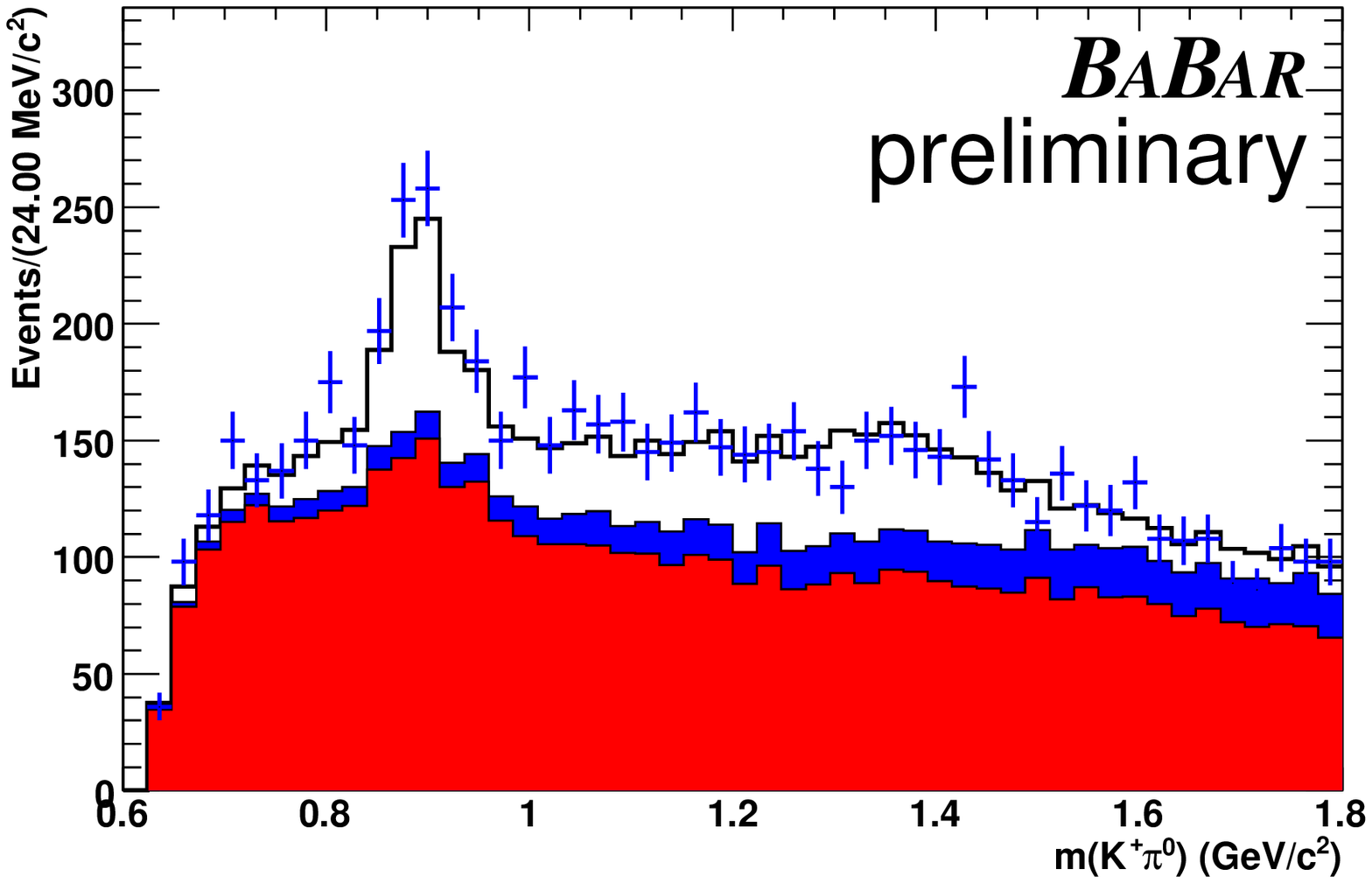,width=7cm}  
\epsfig{file=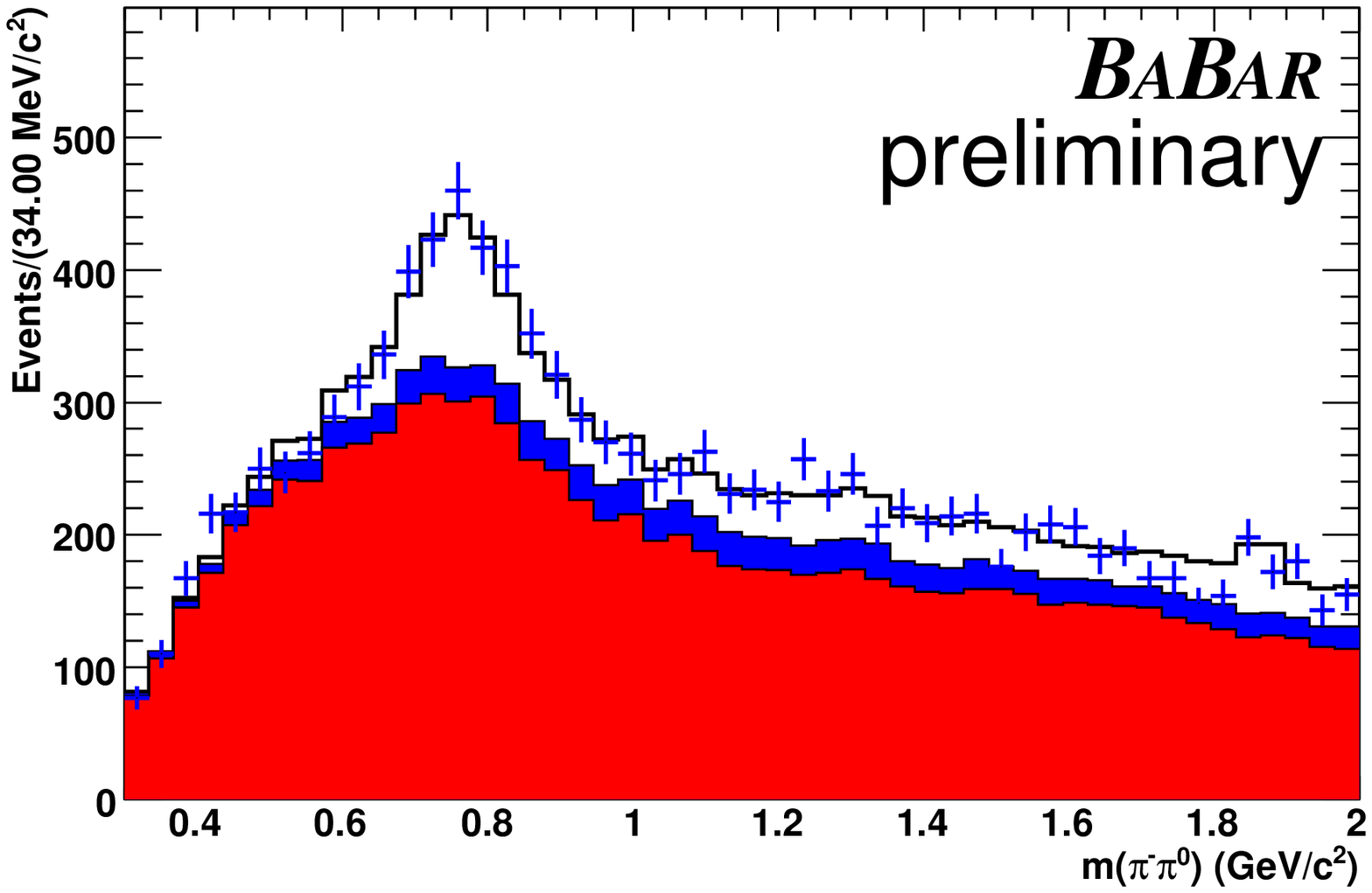,width=7cm}
\caption{\label{fig:DalitzMasseslowmass} Mass distributions for: $m_{\Kp\pim}$ (a), $m_{\Kp\piz}$  (b) and 
$m_{\pim\piz}$ (c). The data are shown as points with error bars. 
The solid histograms show the projection of the fit result. The blue (dark)
and red (gray) shaded areas represent the \B background and continuum, respectively.}
\end{figure}

\begin{figure}[h]
\epsfig{file=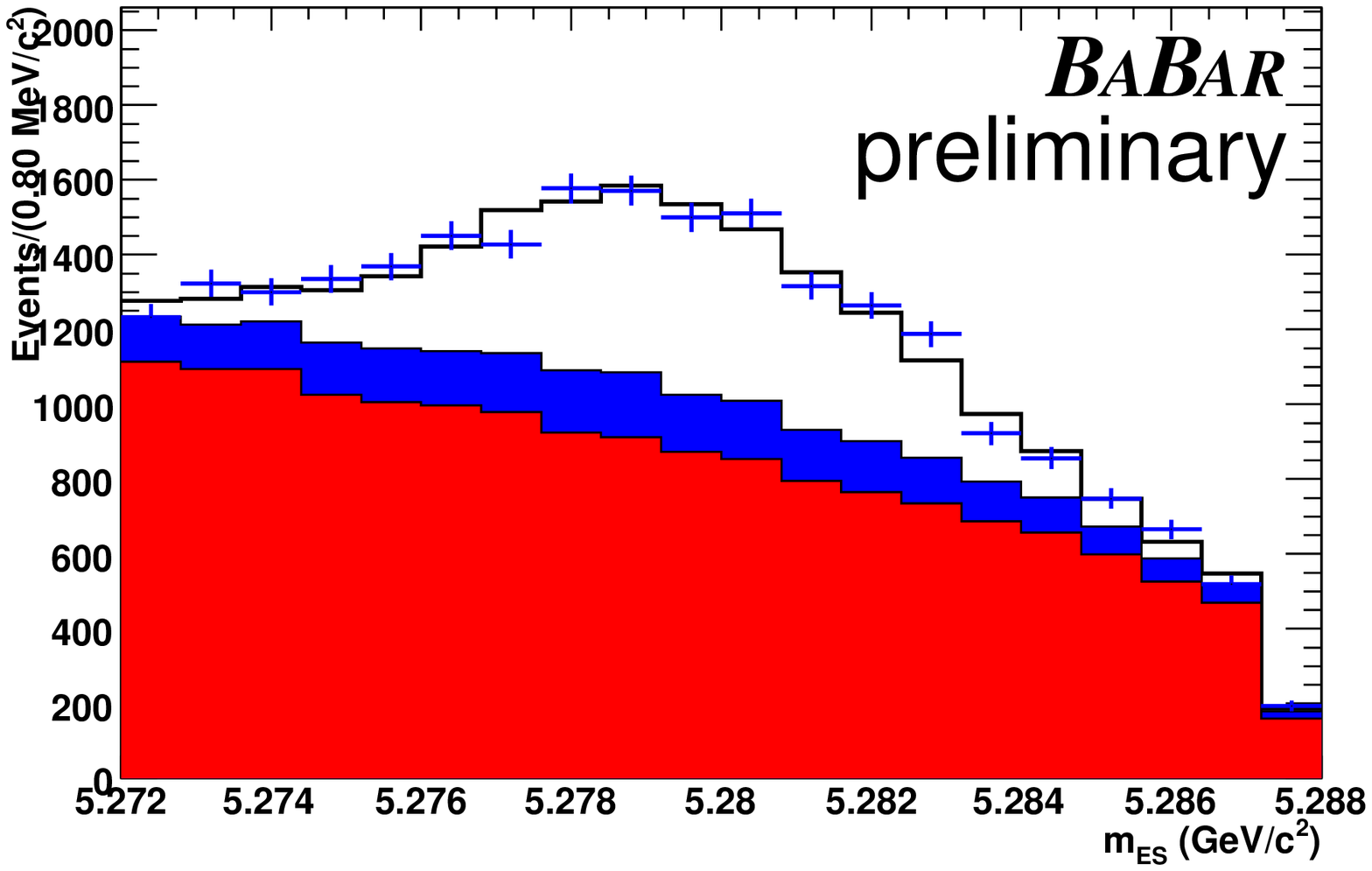,width=7cm}
\epsfig{file=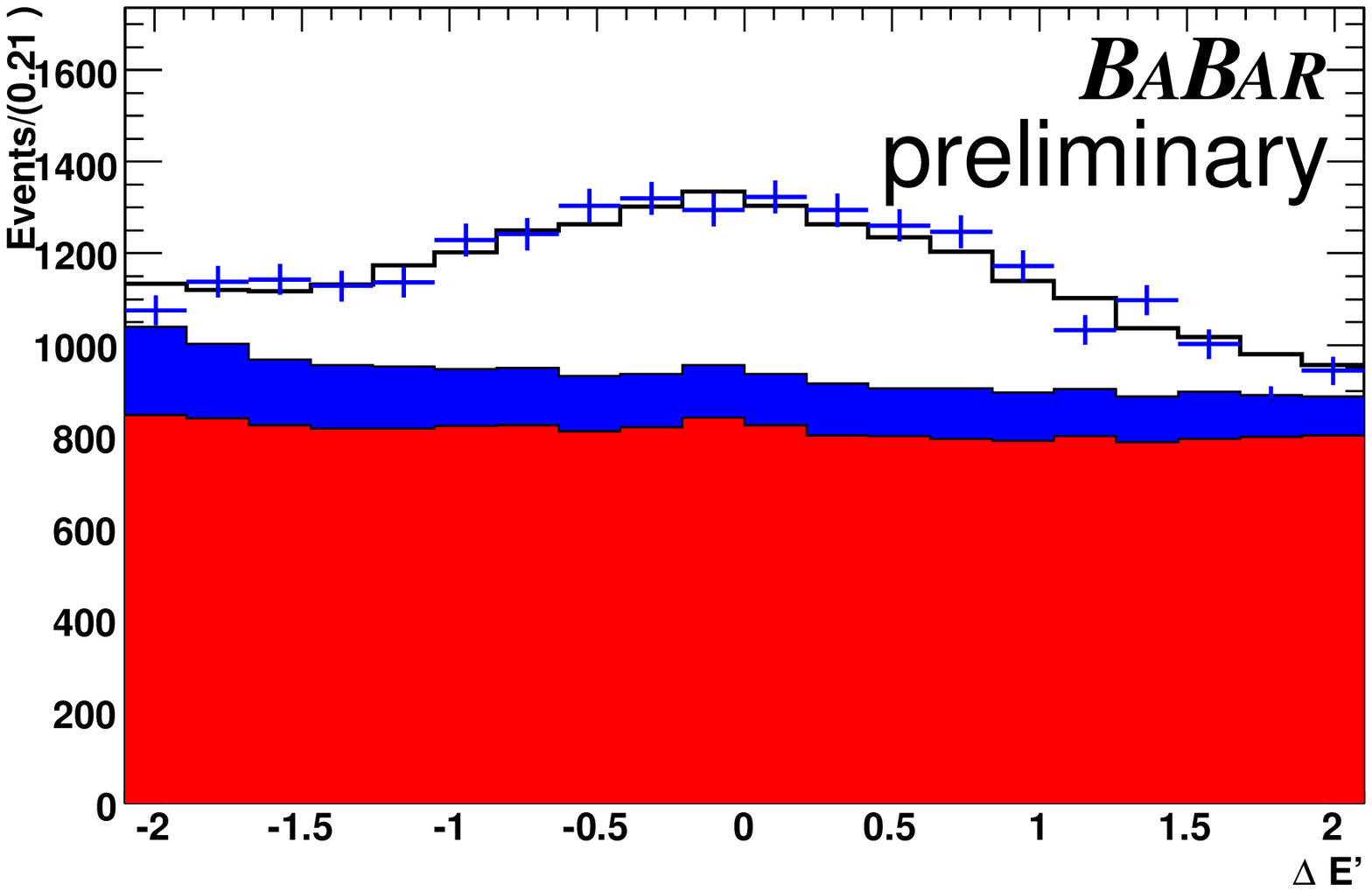,width=7cm}
\epsfig{file=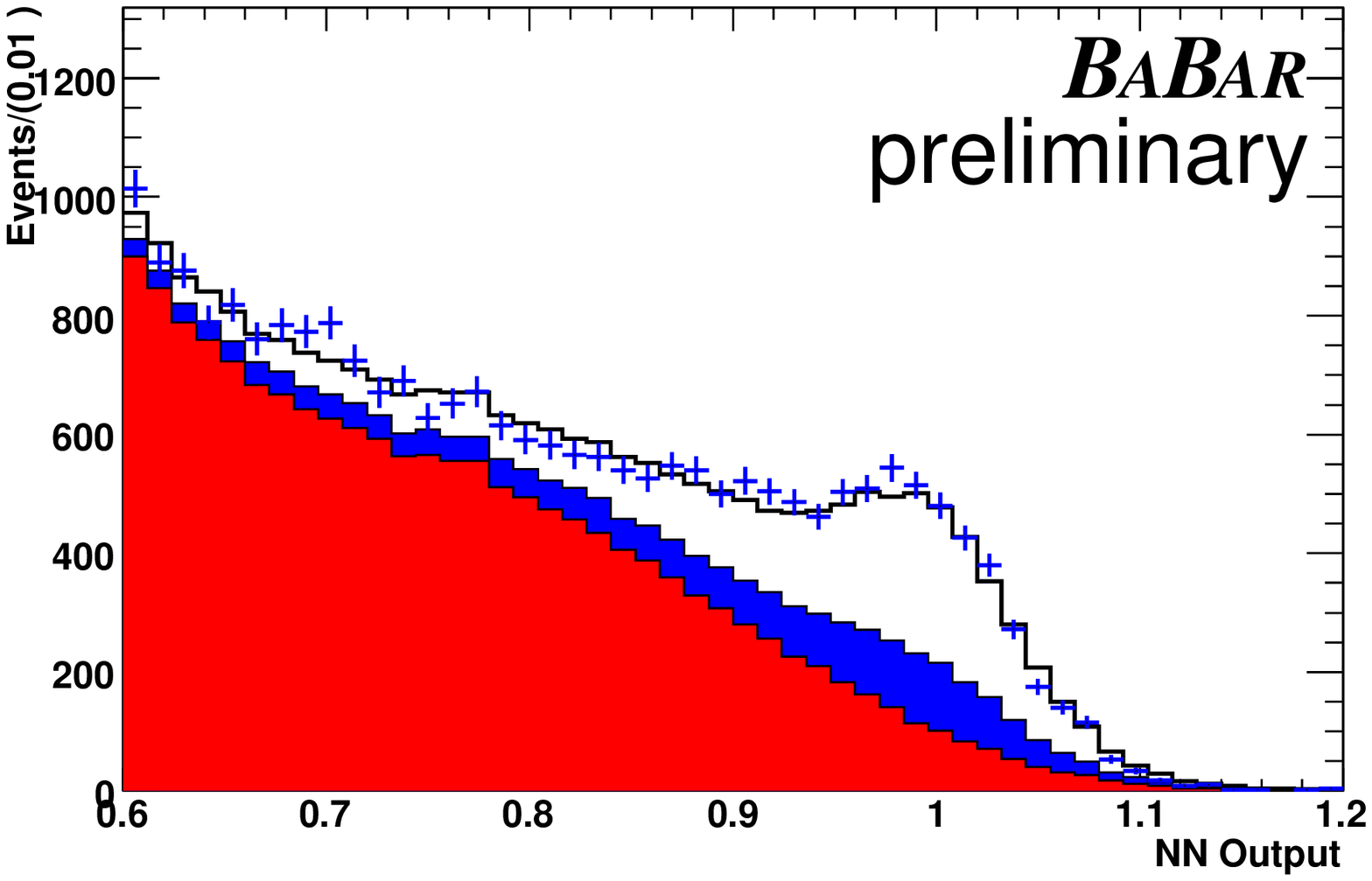,width=7cm} 
\caption{\label{fig:mesdeprimeNN}
(a) \mes, (b) \deprime and (c) NN distributions. The data are shown as points with error bars.  
The solid histograms show the projection of the fit result. The blue (dark)
and red (gray) shaded areas represent the \B background and continuum, respectively.}
\end{figure}

The Dalitz plot mass distributions in an enlargement of the low-mass resonance region (masses below $2.0\ \gevcc$) are shown in \figref{DalitzMasseslowmass} . The $\rho^-$, $\Kstarp$, and $\Kstarz$ are clearly visible in the $m_{\pim\piz}$, $m_{\Kp\piz}$, $m_{\Kp\pim}$ distributions respectively. We calculate a $\chi^2$ of 772 for 644 bins on the Dalitz plot where at least 25 events are guaranteed to exist in each bin. The distributions of the discriminating variables (\mes, \deprime and NN) are shown in~\figref{mesdeprimeNN}. Fitted parameters are given for the four solutions in~\tabref{fourfitsolutions}. We observe that the fit fractions and the $CP$ asymmetries are consistent within less than three standard deviations among the solutions, though the phases differ substantially.

\begin{table*}
\begin{center}
\caption{\label{tab:fourfitsolutions}
Results of the four solutions of the fit. The fractions are the $CP$-averaged isobar fractions ($FF_k$) defined with the $CP$-asymmetries $A_{CP}$ in~\secref{DecayAmplitudes}~[\equaref{PartialFractions}]. The phases $\Phi$ for the $\Bz$ decays and $\overline{\Phi}$ for the $\Bzb$ decays are measured relative to $\Bz(\Bzb)\to\rho^\mp\pipm$. 
The uncertainties are statistical only.}
\setlength{\tabcolsep}{1.2pc}
\begin{tabular}{cccccc}
\hline\hline
Resonance  &  Parameter  &  Solution-I  &  Solution-II  &  Solution-III  &  Solution-IV \\ 
           &  $\Delta (NLL) $   &  0.00  &  3.94 &  7.77  &  10.57 \\ 
\hline
$\rho^-(770) K^+$ &  FF (\%)  & 13.60 $\pm$  1.24 &  13.70 $\pm$  1.25 &  13.20 $\pm$  1.09 &   13.40 $\pm$  1.27 \\
    & $ A_{cp}$  & 0.14 $\pm$  0.06 &  0.17 $\pm$  0.06 &  0.11 $\pm$  0.06 &   0.14 $\pm$  0.06 \\
    & $\overline{\Phi}$  & 0 (fixed) &  0 (fixed) &  0 (fixed)  &   0 (fixed) \\
    & $\Phi$  & 0 (fixed) &  0 (fixed) &  0 (fixed)  &   0 (fixed) \\
\hline
$\rho^-(1450) K^+$ &  FF (\%)  & 4.66 $\pm$  1.42 &  4.13 $\pm$  1.42 &  4.61 $\pm$  1.56 &   4.16 $\pm$  1.61 \\
    & $ A_{cp}$  & 0.14 $\pm$  0.06 &  0.17 $\pm$  0.06 & 0.11 $\pm$  0.06 &  0.14 $\pm$  0.06 \\
    & $\overline{\Phi}$  & 1.63 $\pm$  0.26 &  1.41 $\pm$  0.25 &  1.64 $\pm$  0.28 &   1.39 $\pm$  0.27 \\
    & $\Phi$  & 1.63 $\pm$  0.26 &  1.41 $\pm$  0.25 &  1.64 $\pm$  0.28 &   1.39 $\pm$  0.27 \\
\hline
$\rho^-(1700) K^+$ &  FF (\%)  & 1.16 $\pm$  0.69 &  0.61 $\pm$  0.52 &  0.78 $\pm$  0.64 &   0.30 $\pm$  0.43 \\
    & $ A_{cp}$  & 0.14 $\pm$  0.06 &  0.17 $\pm$  0.06 & 0.11 $\pm$  0.06 &   0.14 $\pm$  0.06 \\
    & $\overline{\Phi}$  & 0.52 $\pm$  0.40 &  0.42 $\pm$  0.51 &  0.38 $\pm$  0.53 &   0.07 $\pm$  0.78 \\
    & $\Phi$  & 0.52 $\pm$  0.40 &  0.42 $\pm$  0.51 &  0.38 $\pm$  0.53 &   0.07 $\pm$  0.78 \\
\hline
$K^{*+}(892)\pi^-$ &  FF (\%)  & 5.52 $\pm$  0.59 &  5.54 $\pm$  0.61 &  5.92 $\pm$  1.21 &   5.88 $\pm$  0.63 \\
    & $ A_{cp}$  & -0.30 $\pm$  0.11 &  -0.30 $\pm$  0.11 &  -0.21 $\pm$  0.11 &   -0.22 $\pm$  0.11 \\
    & $\overline{\Phi}$  & 0.74 $\pm$  0.36 &  0.66 $\pm$  0.36 &  -3.10 $\pm$  0.37 &   3.09 $\pm$  0.36 \\
    & $\Phi$  & 0.37 $\pm$  0.36 &  2.58 $\pm$  0.36 &  0.36 $\pm$  0.36 &   2.61 $\pm$  0.35 \\
\hline
$K^{*0}(892)\pi^0$ &  FF (\%)  & 4.53 $\pm$  0.57 &  4.61 $\pm$  0.57 &  4.63 $\pm$  0.59 &   4.69 $\pm$  0.58 \\
    & $ A_{cp}$  & -0.15 $\pm$  0.12 &  -0.16 $\pm$  0.12 &  -0.15 $\pm$  0.12 &   -0.15 $\pm$  0.12 \\
    & $\overline{\Phi}$  & 0.65 $\pm$  0.29 &  0.58 $\pm$  0.30 &  0.34 $\pm$  0.30 &   0.25 $\pm$  0.30 \\
    & $\Phi$  & -0.00 $\pm$  0.33 &  0.24 $\pm$  0.35 &  -0.03 $\pm$  0.34 &   0.19 $\pm$  0.35 \\
\hline
$(K\pi)^{*+}_0\pi^-$ &  FF (\%)  & 23.60 $\pm$  1.18 &  24.90 $\pm$  1.17 &  24.90 $\pm$  1.19 &   26.10 $\pm$  1.16 \\
    & $ A_{cp}$  & 0.07 $\pm$  0.05 &  0.02 $\pm$  0.05 &  0.11 $\pm$  0.05 &   0.06 $\pm$  0.05 \\
    & $\overline{\Phi}$  & -2.76 $\pm$  0.25 &  -2.84 $\pm$  0.26 &  -0.50 $\pm$  0.32 &   -0.57 $\pm$  0.31 \\
    & $\Phi$  & -2.60 $\pm$  0.30 &  -0.50 $\pm$  0.31 &  -2.60 $\pm$  0.31 &   -0.45 $\pm$  0.31 \\
\hline
$(K\pi)^{*0}_0\pi^0$ &  FF (\%)  & 11.90 $\pm$  1.11 &  17.80 $\pm$  1.24 &  16.60 $\pm$  1.09 &   22.80 $\pm$  1.17 \\
    & $ A_{cp}$  & -0.16 $\pm$  0.09 &  -0.43 $\pm$  0.08 &  0.17 $\pm$  0.07 &  -0.14 $\pm$  0.06 \\
    & $\overline{\Phi}$  & 0.37 $\pm$  0.26 &  0.29 $\pm$  0.27 &  0.20 $\pm$  0.22 &   0.11 $\pm$  0.22 \\
    & $\Phi$  & -0.11 $\pm$  0.27 &  0.26 $\pm$  0.23 &  -0.12 $\pm$  0.28 &   0.26 $\pm$  0.23 \\
\hline
      NR &  FF (\%)  & 5.90 $\pm$  0.93 &  3.98 $\pm$  0.81 &  5.40 $\pm$  1.02 &   3.49 $\pm$  1.09 \\
    & $ A_{cp}$  & 0.07 $\pm$  0.15 &  0.64 $\pm$  0.21 &  -0.04 $\pm$  0.19 &   0.54 $\pm$  0.22 \\
    & $\overline{\Phi}$  & 1.00 $\pm$  0.24 &  0.91 $\pm$  0.24 &  -0.93 $\pm$  0.26 &   -1.04 $\pm$  0.26 \\
    & $\Phi$  & 1.15 $\pm$  0.27 &  -1.22 $\pm$  0.34 &  1.16 $\pm$  0.29 &   -1.21 $\pm$  0.34 \\
\hline
$\overline{D}^0\pi^0$ &  FF (\%)  & 20.90 $\pm$  0.85 &  20.80 $\pm$  0.85 &  20.70 $\pm$  0.93 &   20.60 $\pm$  0.86 \\
\hline
$D^- K^+$ &  FF (\%)  & 0.93 $\pm$  0.23 &  0.96 $\pm$  0.23 &  0.98 $\pm$  0.23 &   1.02 $\pm$  0.24 \\
\hline

\hline\hline
\end{tabular}
\end{center}
\end{table*}
     
\clearpage

\section{SYSTEMATIC UNCERTAINTIES}
\label{sec:Systematics}

\begin{table}[h]
\caption{\label{tab:systmtab} Summary of systematic uncertainties.}
\setlength{\tabcolsep}{1.2pc}
\begin{tabular}{cccccc}
\hline\hline
    Resonance  	          &           &Fit Fraction (\%) &  $A_{cp}$       &  $\Phi$		&  $\overline{\Phi}$ \\ \hline
$\rho^-(770) K^+$ 	& Dalitz Plot Model &  0.500  &  0.003  &  Fixed  &   Fixed  \\ 
                          &  PDF Shape Parameters    &  0.224  &  0.005  &  Fixed  &   Fixed  \\ 
                          & $B$ Backgrounds&  0.071  &  0.001  &  Fixed  &   Fixed  \\ 
                          &  Lineshapes          & 0.229  &  0.003  &  Fixed  &   Fixed  \\ 
                          &  Fit Bias            & 0.015  &  0.004   & Fixed  &   Fixed  \\ 
                          &  {\bf Total}         &  {\bf 0.598}    & {\bf 0.008}      &   {\bf Fixed}       & {\bf Fixed}      \\  \hline
$\rho^-(1450) K^+$ & Dalitz Plot Model &  0.560  &  -  &  0.060  &   0.060  \\ 
                          &  PDF Shape Parameters    &  0.255  & -  &  0.039  &   0.039  \\ 
                          & $B$ Backgrounds&  0.038  &  -  &  0.014  &   0.014  \\ 
                          &  Lineshapes          & 0.232  &  -  &  0.088  &   0.088  \\ 
                          &  Fit Bias            & 0.170  & - &    0.022  &  0.022  \\
                          &  {\bf Total}         & {\bf 0.680}                & -        &   {\bf 0.116}          & {\bf 0.116}      \\  \hline
$\rho^-(1700) K^+$ & Dalitz Plot Model &  0.070  &  -  &  0.077  &   0.077  \\ 
                          &  PDF Shape Parameters    &   0.042  &  -  &  0.038  &   0.038  \\ 
                          & $B$ Backgrounds&  0.009  &  -  &  0.009  &   0.009  \\ 
                          &  Lineshapes          &0.216  &  -  &  0.120  &   0.120  \\ 
                          &  Fit Bias            &0.110   & -   & 0.037  &  0.037\\
                          &  {\bf Total}         & {\bf 0.256}               & -        &   {\bf 0.152}          & {\bf 0.152}        \\  \hline
$K^{*+}(892)\pi^-$ & Dalitz Plot Model &  0.070  &  0.020  &  0.112  &   0.260  \\ 
                          &  PDF Shape Parameters    &   0.154  &  0.026  &  0.021  &   0.080  \\ 
                          &  $\B$ Background     & 0.025  &  0.004  &  0.013  &   0.009  \\ 
                          &  Lineshapes          & 0.030  &  0.006  &  0.061  &   0.147  \\ 
                          &  Fit Bias            &   0.020  &0.004 &   0.047 &   0.038 \\
                          &  {\bf Total}         & {\bf 0.175}              & {\bf 0.034}      &   {\bf 0.138}   & {\bf 0.312}        \\  \hline
$K^{*0}(892)\pi^0$ & Dalitz Plot Model &  0.200  &  0.010  &  0.065  &   0.065  \\ 
                          &  PDF Shape Parameters    &   0.128  &  0.005  &  0.020  &   0.031  \\ 
                          &  $\B$ Background         & 0.028  &  0.002  &  0.015  &   0.012  \\ 
                          &  Lineshapes          &0.039  &  0.004  &  0.048  &   0.065  \\ 
                          &  Fit Bias            &  0.087  &0.005 &    0.046 &   0.012  \\
                          &  {\bf Total}         & {\bf 0.257}    & {\bf 0.022}        &   {\bf 0.096}        & {\bf 0.098}       \\  \hline
$(K\pi)^{*+}_0$ & Dalitz Plot Model &  1.200  &  0.009  &  0.050  &   0.190  \\ 
                          &  PDF Shape Parameters    &    1.166  &  0.007  &  0.022  &   0.079  \\ 
                          &  $\B$ Background         &0.071  &  0.001  &  0.012  &   0.015  \\ 
                          &  Lineshapes          &0.166  &  0.004  &  0.047  &   0.121  \\ 
                          &  Fit Bias            & 0.260 &   0.006    & 0.041&   0.030   \\
                          &  {\bf Total}         & {\bf 1.703}  & {\bf 0.014}        &   {\bf 0.084}         & {\bf 0.218}       \\  \hline
$(K\pi)^{*0}_0$ & Dalitz Plot Model &  1.400  &  0.033  &  0.370  &   0.225  \\ 
                          &  PDF Shape Parameters    &   0.173  &  0.016  &  0.056  &   0.032  \\ 
                          &  $\B$ Background         &0.071  &  0.005  &  0.025  &   0.016  \\ 
                          &  Lineshapes          &0.224  &  0.017  &  0.071  &   0.071  \\ 
                          &  Fit Bias            & 0.300   & 0.003 &   0.044 &   0.009 \\
                          &  {\bf Total}         & {\bf 1.461}    & {\bf 0.041}        &   {\bf 0.384}         & {\bf 0.239}        \\  \hline
      NR                  & Dalitz Plot Model &  0.240  &  0.017  &  0.119  &   0.110  \\ 
                          &  PDF Shape Parameters    &   0.744  &  0.035  &  0.106  &   0.060  \\ 
                          &  $\B$ Background         & 0.042  &  0.003  &  0.014  &   0.014  \\ 
                          &  Lineshapes          &  0.134  &  0.019  &  0.067  &   0.117  \\ 
                          &  Fit Bias            &  0.120  &  0.003  &  0.009  &  0.034  \\
                          &  {\bf Total}         & {\bf 0.803}    & {\bf 0.044}        &   {\bf 0.174}        &  {\bf 0.175}      \\  \hline
$\overline{D}^0\pi^0$	  & Dalitz Plot Model &  0.500  &  -  &  -  &   -  \\ 
                          &  PDF Shape Parameters    &  2.606  &  -  &  -  &   -  \\ 
                          &  $\B$ Background         & 0.150  &   -  &   -  &   -  \\ 
                          &  Lineshapes          & 0.071  &   -  &  -  &   -  \\ 
                          &  Fit Bias            & 0.073  &   -  &  -  &   -  \\ 
                          &  {\bf Total}         & {\bf 2.660}               & -  &- &-\\ \hline
$D^- K^+$ 	          & Dalitz Plot Model &  0.010  &  -  &  -  &   -  \\ 
                          &  PDF Shape Parameters    &  0.015  &  -  &  -  &   -  \\ 
                          &  $\B$ Background         &0.002  &   -  &   -  &   -  \\ 
                          &  Lineshapes          &0.013  &   -  &  -  &   -  \\ 
                          &  Fit Bias            &0.001  &   -  &  -  &   -  \\ 
                          &  {\bf Total}         & {\bf 0.022}  & -  &- &-\\ \hline\hline
\end{tabular}
\end{table}

Variations around the nominal fit are tried to study the dominant systematic effects, summarized in~\tabref{systmtab}.
For each parameter of interest  ($FF$, $A_{CP}$, $\Phi$), the positive (negative) deviations from each effect are summed in quadrature to obtain total upward (downward) systematic errors $\delta_+$ ($\delta_-$). Systematic effects are studied by varying the number of resonances contributing to the signal model, the lineshape parameters of the resonances in the signal model, the yields of the nominally fixed B-backgrounds, and the shape of the continuum Dalitz PDF. The intrinsic bias of the fit as measured in MC studies, is also included as a source of systematic error. 
\par

\begin{itemize}
\item To estimate the contribution of other resonances, we fit the on resonance data with extended signal models including one {\it extra}-resonance in addition to those in the nominal signal model. The $\Kstarz_2(1430)\piz,\ \Kstarp_2(1430)\pim$, $\Kstarz(1680)\piz$ and $\Kstarp(1680)\pim$ have been added to the nominal model, and fits with these resonances show some improvement in likelihood. In the best solutions obtained with these extended models the addition resonances did not significantly interfere with those in the nominal model. The variations of the physical parameters due to additional resonances are recorded as {\it Dalitz Plot Model} uncertainties in~\tabref{systmtab}. 
\par

\item The variations of the physical parameters of the resonances in the nominal signal model are recorded as {\it Lineshape} systematic uncertainties.\par

\item Variations of the PDF shape parameters are recorded as {\it PDF Shape Parameter } systematic uncertainties. Specifically, mismodeling of the continuum square Dalitz plot PDF~(Section V.B) is studied by recreating the PDF with numerous smoothing parameters and varying the amount of B-background subtracted from the \mes sideband by 50\%. A small difference in the shape of the TM-signal NN distribution between data and MC is also studied.\par 
 
\item To estimate the {\it Fit Bias} uncertainties inherent in our fit technique, we record the fitted biases and spreads in fits performed on large Monte Carlo samples with both signal and background events generated with their nominal PDFs. \par

\item Each of the nominally fixed B-background yields is allowed to vary freely in a series of fits to data. The variations of B-background yields are recorded as {\it B-background} systematic uncertainties. \par

\end{itemize}

\section{SUMMARY}
\label{sec:Summary}
We have performed an amplitude analysis of the $\Bz\to\Kp\pim\piz$ decay. We have measured the $CP$-averaged fit fractions, $CP$-asymmetries
and phases of the decay precesses to the intermediate states with $\rho^-(770)\Kp$, $\rho^-(1450)\Kp$, $\rho^-(1700)\Kp$, $K^*(892)^{+,0}\pi^{-,0}$, $(K\pi)^{*+,0}_0\pi^{-,0}$. We find a satisfactory solution that provides a significant constraint on the phases of the resonances. For this solution, the CP asymmetries are consistent with zero in all quasi two-body channels. Three further solutions were found, though all had an NLL worse by 3.9 units, or more. Additionally, we measure the $CP$-averaged fit fractions for the decays $\Bz\to\overline{D}^0\piz\to\Kp\pim\piz$ and $\Bz\to D^-\Kp\to\Kp\pim\piz$.

\clearpage
\section{Acknowledgments}
\label{sec:acknowledgments}
We are grateful for the 
extraordinary contributions of our \pep2\ colleagues in
achieving the excellent luminosity and machine conditions
that have made this work possible.
The success of this project also relies critically on the 
expertise and dedication of the computing organizations that 
support \babar.
The collaborating institutions wish to thank 
SLAC for its support and the kind hospitality extended to them. 
This work is supported by the
US Department of Energy
and National Science Foundation, the
Natural Sciences and Engineering Research Council (Canada),
the Commissariat \`a l'Energie Atomique and
Institut National de Physique Nucl\'eaire et de Physique des Particules
(France), the
Bundesministerium f\"ur Bildung und Forschung and
Deutsche Forschungsgemeinschaft
(Germany), the
Istituto Nazionale di Fisica Nucleare (Italy),
the Foundation for Fundamental Research on Matter (The Netherlands),
the Research Council of Norway, the
Ministry of Education and Science of the Russian Federation, 
Ministerio de Educaci\'on y Ciencia (Spain), and the
Science and Technology Facilities Council (United Kingdom).
Individuals have received support from 
the Marie-Curie IEF program (European Union) and
the A. P. Sloan Foundation.


\begin{thebibliography}{99}
\bibitem{Cmodeimplied}  Throughout the paper, whenever a mode is given, the charge conjugate is also implied.
\end{thebibliography}
\end{document}